\documentclass[twocolumn,trackchanges]{aastex701}

\usepackage{amsmath}
\usepackage{caption}

\begin{document}

\title{The Ultrafast Line-Driven Wind from the Double-Degenerate Merger Remnant WD J005311}

\author[orcid=0009-0000-9809-1518,gname=Kazuma, sname='Kato']{Kazuma Kato}
\affiliation{Astronomical Institute, Tohoku University, Sendai, Miyagi 980-8578, Japan}
\email[show]{kazuma.kato@astr.tohoku.ac.jp}

\author[orcid=0000-0003-4299-8799,gname=Kazumi, sname='Kashiyama']{Kazumi Kashiyama} 
\affiliation{Astronomical Institute, Tohoku University, Sendai, Miyagi 980-8578, Japan}
\email{kashiyama@astr.tohoku.ac.jp}

\begin{abstract}
We systematically construct steady-state wind solutions for double-degenerate merger remnants consisting of a degenerate oxygen-neon core and an optically thick expanding envelope powered by carbon-shell burning, 
with winds accelerated by radiation pressure including line driving. For a given envelope composition, each solution is characterized by two eigenvalues: the degenerate core mass, $M_{\rm WD}$, and the total envelope and wind mass,
$\Delta M$. We find that wind solutions exist for $M_{\rm WD} \gtrsim 1.0\,M_\odot$. For a given $M_{\rm WD}$, the solutions transition with decreasing $\Delta M$ from slow, continuum-driven winds to ultrafast, line-driven winds, 
forming a sequence that can be interpreted as the temporal evolution of a merger remnant. We apply these solutions to WD J005311, a Galactic double-degenerate merger-remnant candidate with an ultrafast wind of $v_{\rm w} \sim 0.05\,c$. Its luminosity, 
effective temperature, and mass-loss rate are consistently reproduced with $M_{\rm WD} \sim 1.15\mbox{--}1.25\,M_\odot$ and $\Delta M \sim (1.8\mbox{--}5.7)\times 10^{-3}\,M_\odot$. 
Combining our evolutionary analysis with observations of the surrounding nebula Pa 30 and historical records of SN 1181, 
we argue that WD J005311 began launching a continuum-driven wind $400\mbox{--}650$ yr after the putative merger and transitioned to the ultrafast-wind phase approximately $100$ yr ago. 
This phase is expected to continue for another $\sim 1000$ yr, during which the WD will remain below the Chandrasekhar mass and avoid collapse into a neutron star. 
Before the continuum-driven wind phase, the remnant may have passed through a more bloated, hydrostatic giant phase with a slower but more massive wind. Future observations of the wind nebula could test this scenario by revealing signatures of interactions between the slow, 
massive wind and the ultrafast wind.
\end{abstract}


\section{Introduction} 
Binary white dwarfs (WDs) are the most abundant class of binary compact objects.
Some systems have sufficiently small orbital separations that they
undergo merger driven by gravitational-wave emission,
leading to a variety of outcomes depending on their
masses and compositions.
Since the 1980s, particular attention has been paid to systems with total masses exceeding the Chandrasekhar mass, as they may produce a Type Ia supernova
\citep{1984ApJS...54..335I,1984ApJ...277..355W},
which serves as an important distance indicator in cosmology,
or undergo collapse to form a neutron star
\citep{1985A&A...150L..21S,1985ApJ...297..531N}.
Although these systems are of great astrophysical importance,
there is still no consensus regarding the conditions that determine their ultimate fate,
partly because merger remnants are difficult to observe directly.

In 2019, \citet{2019Natur.569..684G} reported the discovery of an extremely hot WD
(WD J005311) located at the center of the hydrogen and helium deficient
mid-infrared nebula Pa30.
Spectroscopic observations revealed that the atmosphere of this WD is also
deficient in hydrogen and helium, and that it has an exceptionally high
effective temperature of
$T_{\rm eff}\approx 2\times10^{5}\ {\rm K}$.
Furthermore, using distance measurements from Gaia DR3
\citep{2021AJ....161..147B},
its luminosity was estimated to be
$\approx3.0-6.0\times10^{4}\,L_{\odot}$,
which is close to the Eddington luminosity of a one solar mass object
\citep{2019Natur.569..684G,2023ApJ...944..120L}.
The high luminosity and effective temperature may be attributed to energy
generation from carbon shell burning near the surface of the oxygen-neon
(ONe) degenerate core, and these properties are indeed consistent with
theoretical models of carbon-oxygen (CO) WD and ONe WD merger remnants
in the quasi-static phase \citep{2016MNRAS.463.3461S, 2023MNRAS.525.6295W}.
The single-degenerate scenario involving accretion from a He star has also
been proposed, although the lack of a surviving He-star companion disfavors
this scenario \citep{2026arXiv260719705U}.
Therefore, this object has been regarded as a candidate remnant of a merger
between a CO WD and an ONe WD.

Pa30 has also been studied in detail.
These observations have suggested that the total mass of the nebula is approximately $0.1\mbox{--}0.5\, M_{\odot}$ and that its expansion velocity is about $1,000 {\rm km~s^{-1}}$
\citep{2020A&A...644L...8O, 2024ApJ...969..116K, 2021ApJ...918L..33R}.
The corresponding kinetic energy is estimated to be of order $10^{48}\ {\rm erg}$,
which is consistent with that of Type Iax supernovae, a class of explosions that may be associated with mergers involving CO and/or ONe WDs.
These results therefore support the hypothesis that WD J005311 is a WD merger remnant.
Moreover, this object was identified as the remnant of SN 1181, a historical supernova recorded as a guest star,
based on the ejection age inferred from the expansion velocity, spatial distribution, and direction of the ejecta \citep{2021ApJ...918L..33R}.
As a result, WD J005311 is regarded as a unique candidate WD merger remnant
for which the time since the merger event can be constrained with unusual precision.

The merger remnant of an ONe WD and a CO WD is expected to consist of
a massive degenerate core originating from the ONe WD, surrounded by
an envelope composed of material from the disrupted CO WD.
The envelope can have two distinct structural phases.
In one phase, the envelope forms an extended configuration
in hydrostatic equilibrium
\citep{2012ApJ...748...35S, 2016MNRAS.463.3461S, 2017ApJ...850..127B,
2021ApJ...906...53S, 2026arXiv260311190P}.
We refer to this phase as the ``hydrostatic giant phase.''
In the other phase, the envelope forms an optically thick,
steady wind rather than remaining in hydrostatic equilibrium
(e.g., \citealt{1994ApJ...437..802K}).
We refer to this phase as the ``optically thick wind phase.''

A particularly striking feature of WD J005311 is its ultrafast wind.
From the widths of the O\,IV emission lines,
the wind velocity has been estimated to be as high as
$16,000\ {\rm km~s^{-1}}$
\citep{2019Natur.569..684G,2023ApJ...944..120L,2026arXiv260520360T}.
Given the observed photospheric radius of approximately $0.1\,R_\odot$, 
the wind velocity is significantly higher than the escape velocity at this radius, 
indicating that an efficient acceleration mechanism operates near the photosphere and suggesting 
that the object is likely in an optically thick wind phase.

\citet{2019ApJ...887...39K} and \citet{2024ApJ...963...26Z}
constructed optically thick magnetocentrifugal wind models
to explain the observed properties of WD J005311,
particularly its ultrafast wind velocity.
However, if this object originated from a WD binary merger, 
it likely underwent a hydrostatic giant phase during 
which substantial angular momentum loss would have occurred. 
Such angular momentum loss is expected to significantly reduce the stellar rotation rate,
making it difficult to sustain the rapid rotation assumed in these models \citep{2021ApJ...906...53S}.
Indeed, recent observations suggest that the rotation period
of WD J005311 is longer than assumed in the magnetocentrifugal
models \citep{2026arXiv260520360T}. 
In addition, the strong magnetic fields
and rapid rotation required by the magnetocentrifugal scenario may be difficult to reconcile with the nearly spherical morphology of the observed nebula.
Taken together, these arguments suggest that the
magnetocentrifugal mechanism may not be the dominant driver of the wind from WD J005311.

Recent spectroscopic observations have also revealed temporal variability
in the spectrum of WD J005311, suggesting that the wind has a
clumpy structure \citep{2026arXiv260520360T}. Such a feature is
consistent with a line-driven wind, since radiative line driving
is expected to generate density inhomogeneities through instabilities to velocity perturbations
which grow on scales smaller than the Sobolev length \citep{1984ApJ...284..337O, 1999ApJ...514..909L}.
These observational findings motivate us to
explore a radiation-driven wind model, including radiative line driving,
as an alternative to the magnetocentrifugal scenario.

\begin{figure*}[t]
\centering
\includegraphics[width=0.9\textwidth]{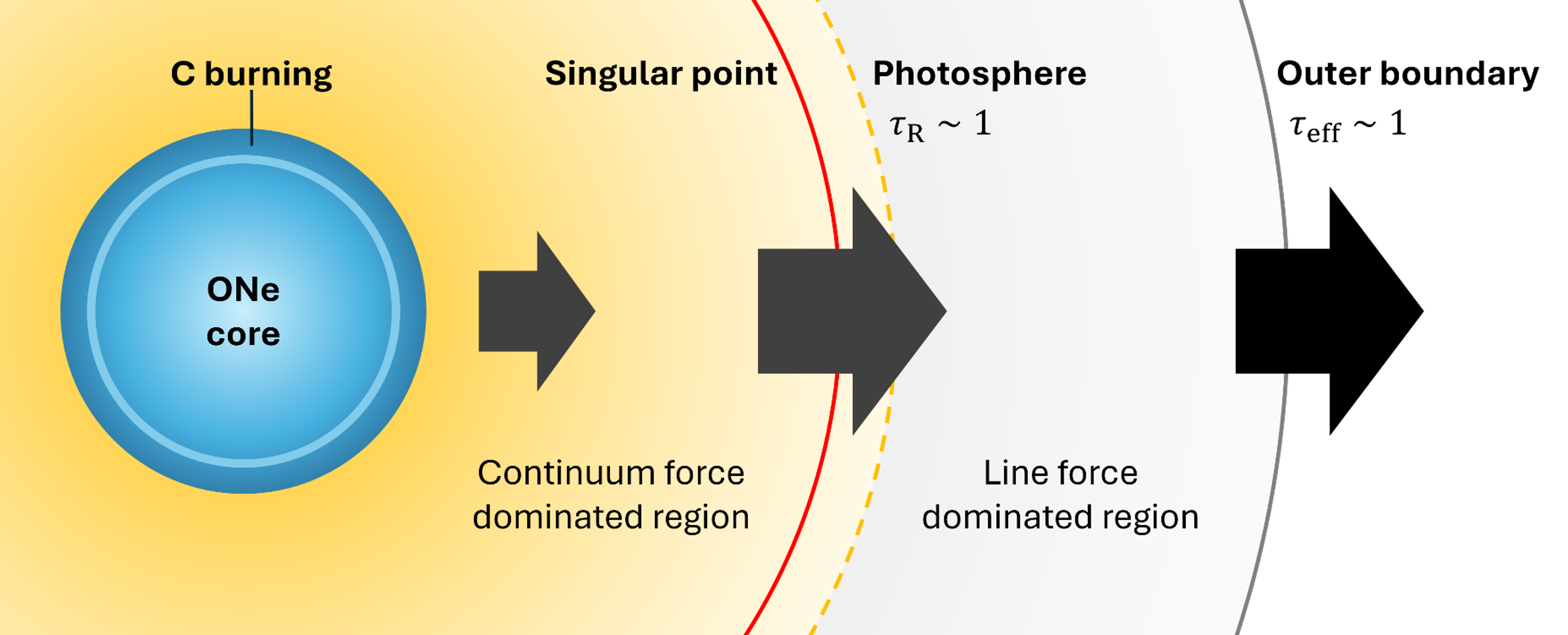}
\caption{
Schematic picture of the line-driven wind model.
\label{fg:modelcartoon}}
\end{figure*}

Here, we propose a model of an optically thick wind in which line force
becomes important near the photosphere, which we refer to as the
``line-driven wind'' model hereafter (Fig.~\ref{fg:modelcartoon}).
The model consists of two components: a degenerate core and a wind
region extending above its surface.
Carbon burning occurs near the surface of the ONe degenerate core
and supplies radiative energy to the overlying wind.
In the relatively inner wind region, where the optical depth defined
by the Rosseland mean opacity is larger than unity, the wind is
predominantly accelerated by continuum radiation.
Farther out, radiative acceleration due to spectral lines becomes
important.

In this paper, we systematically investigate the optically thick wind
phase of WD merger remnants using the line-driven wind model.
Based on these results, we propose a wind model that accounts for the observations of WD J005311.
We also discuss the evolutionary history and future evolution of WD J005311 based on the sequence of the line-driven wind solutions.

In Sec.~\ref{sec:windmodel}, we describe the equations governing steady-state winds from WDs and the adopted boundary conditions.
In Sec.~\ref{sec:calmeth}, we describe the numerical method used to obtain the wind solutions and the parameter space explored in this study.
In Sec.~\ref{sec:result}, we present the wind solutions and compare them with the observations.
In Sec.~\ref{sec:discussion}, we discuss the implications of our results for the physical properties, evolutionary history, and ultimate fate of WD J005311.

\section{The line-driven wind model} \label{sec:windmodel}
First, we present the basic equations of the line-driven wind model, in Sec.~\ref{ssec:basiceq}.
Next, we explain how the line force is calculated in Sec.~\ref{ssec:lineforce}.
We then discuss the boundary conditions in Sec.~\ref{ssec:boundary}.

\subsection{Basic Equations} \label{ssec:basiceq}
We assume that the wind is steady and spherical.
To formulate the equation of motion, we consider radiative forces
from both continuum and line driving.
The radiative acceleration per unit mass due to continuum processes
is written as
\begin{equation}
    f^{\rm con}
    =
    \frac{\kappa_{\rm R} L}{4\pi c r^2},
\end{equation}where $r$ is the radial distance from the center of the degenerate core,
$L$ is the luminosity, $\kappa_{\rm R}$ is the Rosseland mean opacity
calculated using the Los Alamos OPLIB opacity tables
\citep{2016ApJ...817..116C}, and $c$ is the speed of light.
On the other hand, the radiative acceleration per unit mass due to
line absorption is written as
\begin{equation}
    f^{\rm line}
    =
    \frac{\kappa_{\rm es,f} M(t)L}{4\pi r^2 c}.
\end{equation}Here, $\kappa_{\rm es,f}$ represents the electron-scattering opacity
in the fully ionized limit, and $M(t)$ is the dimensionless
line-force multiplier defined in Sec.~\ref{ssec:lineforce}.
Including both continuum and line contributions,
the total radiative acceleration is written as
\begin{equation}
    f^{\rm rad}
    =
    \kappa_{\rm R}
    \left(
    1+
    \frac{\kappa_{\rm es,f}}{\kappa_{\rm R}}M(t)
    \right)
    \frac{L}{4\pi r^2 c}.
\end{equation}The first term becomes important in continuum force dominated regions,
whereas the second term becomes important in line force dominated regions.
This formulation captures the essential effects
of both continuum and line driving throughout the wind.
We therefore define the effective opacity $\kappa_{\rm eff}$ as
\begin{equation}
    \kappa_{\rm eff}
    \equiv
    \kappa_{\rm R}
    \left(
    1+
    \frac{\kappa_{\rm es,f}}{\kappa_{\rm R}}M(t)
    \right),
\end{equation}so that the total radiative acceleration can be written as
\begin{equation}
    f^{\rm rad}
    =
    \frac{\kappa_{\rm eff} L}{4\pi r^2 c}.
\end{equation}Although approximate, this effective opacity provides a good
description of the average strength of photon--matter interactions
throughout the wind region.
Therefore, hereafter, we formulate photon--matter interactions in the
wind using this effective opacity $\kappa_{\rm eff}$.

Using $\kappa_{\rm eff}$, the equation of motion can be written as
\begin{equation}
    v\frac{dv}{dr}
    +
    \frac{1}{\rho}\frac{dp}{dr}
    +
    \frac{GM_{\rm WD}}{r^2}
    -
    \frac{\kappa_{\rm eff}L}{4\pi r^2 c}
    =
    0, \label{eq:Euler}
\end{equation}where 
$v$ is the wind velocity, $\rho$ is the mass density,
$p$ is the gas pressure, $M_{\rm WD}$ is the mass of the
degenerate core, and $G$ is the gravitational constant.
We assume that the mass of the wind region is negligible compared to the degenerate core mass;
therefore, $M_{\rm WD}$ is treated as constant.
In Eq.~\ref{eq:Euler},
the second, third, and fourth terms represent the gas-pressure gradient,
gravitational acceleration,
and radiative acceleration, respectively.

The pressure is assumed to obey the ideal-gas equation of state,
\begin{equation}
    p
    =
    \frac{k_{\rm B}\rho T}{\mu m_{\rm H}},
\end{equation}where $T$ is the temperature,
$\mu$ is the mean molecular weight,
$m_{\rm H}$ is the hydrogen mass,
and $k_{\rm B}$ is the Boltzmann constant.
The mean molecular weight $\mu$
\footnote{In our model, radiation acceleration is the dominant source of fluid acceleration,
so assuming a constant mean molecular weight does not affect the wind solutions.}
 does not vary significantly throughout the wind, 
and we therefore adopt the value for a fully ionized gas in all calculations.

For energy transport in regions where the optical depth
defined by $\kappa_{\rm eff}$ is much larger than unity,
the diffusion approximation can be applied:
\begin{equation}
\frac{dT}{dr}
=
-
\frac{
3\kappa_{\rm eff}\rho L
}{
16\pi a c T^3 r^2
}. \label{eq:effdiffusion}
\end{equation}where $a$ is the radiation constant.

Assuming stationarity, the continuity equation and the energy
conservation equation can be written in integral forms as
\begin{equation}
    4\pi r^2 \rho v
    =
    \dot{M}, \label{eq:cont}
\end{equation}and
\begin{equation}
    L
    +
    \dot{M}
    \left(
    w
    +
    \frac{v^2}{2}
    -
    \frac{GM_{\rm WD}}{r}
    \right)
    =
    \Lambda_{\rm tot}, \label{eq:energycons}
\end{equation}where $\dot{M}$ is the mass-loss rate, which is constant throughout
the flow, $w$ is the specific enthalpy of the gas and radiation,
and $\Lambda_{\rm tot}$ is the total energy release rate.
The enthalpy is given by
\begin{equation}
    w
    =
    \frac{5}{2}
    \frac{k_{\rm B}T}{\mu m_{\rm H}}
    +
    \frac{4aT^4}{3\rho}.
\end{equation}

By solving the above equations,
we obtain the radial profiles of $v$, $T$, $\rho$, and $L$.

\subsection{The Line Force Multiplier} \label{ssec:lineforce}
In regions outside the photosphere defined by the Rosseland mean
opacity, radiative acceleration due to spectral lines becomes important.
Because neighboring fluid elements have different velocities,
the resonance frequencies of bound--bound transitions are Doppler shifted,
allowing photons to continuously transfer momentum to the outflowing gas.
We describe here the calculation of the line force \citep{1975ApJ...195..157C, 1986PhT....39f..90M, 1999isw..book.....L}.

Under the Sobolev approximation,
the radiative acceleration associated with a single
bound--bound transition labeled by $s$ is given by \citep{1974MNRAS.169..279C}
\begin{equation}
    f^{\rm line}_{s}
    =
    \frac{
    \alpha_s \Delta \nu_{\rm D} F(\nu_s)
    }{
    c\rho
    }
    \frac{1-e^{-\tau_s}}{\tau_s},
\end{equation}where $\alpha_s$ is the line absorption coefficient,
$\nu_s$ is the line-center frequency,
$\Delta \nu_{\rm D}$ is the Doppler width due to thermal motion and
$F(\nu_s)$ is the spectral flux evaluated at $\nu_s$.
The quantity $\tau_l$ denotes the Sobolev optical depth,
\begin{equation}
    \tau_s
    =
    \alpha_s v_{\rm th}
    \left|\frac{dv}{dr}\right|^{-1},
\end{equation}where $v_{\rm th}$ is the thermal velocity.
A larger velocity gradient produces a larger Doppler shift
between neighboring fluid elements in the observer frame,
thereby reducing the Sobolev optical depth and enhancing
the line acceleration.

In order to evaluate the strength of the line force,
the contributions from all bound--bound transitions must be summed.
Summing over all spectral lines,
the total radiative acceleration due to lines can be expressed as
\begin{equation}
    f^{\rm line}
    =
    \frac{\kappa_{\rm es,f} F}{c} M(t),
\end{equation}where $M(t)$ is the dimensionless line-force multiplier,
and $t$ is the optical-depth parameter defined by
\begin{equation}
    t
    =
    \kappa_{\rm es,f}\rho c_{\rm s}
    \left|
    \frac{dv}{dr}
    \right|^{-1}.
\end{equation}where $\kappa_{\rm es,f}$ represents the electron-scattering opacity
in the fully ionized limit.
In this study,
the line-force multiplier is defined relative to
the electron-scattering opacity in the fully ionized limit.

Following \citet{2021AAS...23711602L},
the line-force multiplier can be written as
\begin{equation}
    M\left(t\right)
    =
    \eta \sum_{s} q_{s}\tilde{W}_s
    \left(
    \frac{1-e^{-\tau_{s}}}{\tau_s}
    \right), \label{eq:forcemult}
\end{equation} with
\begin{equation}
    q_s
    =
    \frac{3}{8}
    \frac{\lambda^{\rm el}_{i,jk}}{r_{\rm e}}
    f_{jk}
    \frac{n^{\rm el}_{i,j}}{n_{\rm e,f}}
    \left(
    1-e^{-\frac{h\nu^{\rm el}_{i,jk}}{k_{\rm B}T}}
    \right),
\end{equation}
\begin{equation}
    \tilde{W}_s
    =
    \frac{\nu^{\rm el}_{i,jk}F_{\nu^{\rm el}_{i,jk}}}{F},
\end{equation}and
\begin{equation}
    \tau_s
    =
    \frac{c}{c_s}q_s t.
\end{equation}Here, $l$ labels a bound--bound transition,
while $j$ and $k$ denote the lower and upper levels of the transition,
respectively.
The quantities $\lambda^{\rm el}_{i,jk}$ and $\nu^{\rm el}_{i,jk}$
are the wavelength and frequency of the transition,
$f^{\rm el}_{i,jk}$ is the oscillator strength,
and $F_{\nu^{\rm el}_{i,jk}}$ is the spectral flux evaluated at
$\nu^{\rm el}_{i,jk}$.
The quantity $c_s$ is the isothermal sound speed,
$\eta$ is the geometrical finite-disk factor,
$r_{\rm e}$ is the classical electron radius and $h$ is the Planck constant.
For simplicity, we assume the finite-disk factor to be unity.
Finally, $n^{\rm el}_{i,j}$ denotes the number density of particles
of species ``${\rm el}$'' in ionization stage $i$ and excitation level $j$.
Since this quantity is determined by the local chemical composition,
temperature, and density, the line-force multiplier also depends on
these local properties.
Note that $n_{\rm e,f}$ denotes the electron number density
in the fully ionized limit,
since the line-force multiplier is defined relative to
the electron-scattering opacity for fully ionized gas.

We assume that the chemical composition is spatially uniform
throughout the wind.
Even in this case, however, the density and temperature vary throughout the wind.
Therefore, in order to calculate the contribution of the line force
accurately, it is necessary to evaluate the contributions from all
bound--bound transitions according to the local temperature and density,
and to determine the corresponding form of $M(t)$.

However, because the number of spectral lines associated with heavy
elements is extremely large, calculating $M(t)$ directly at every point
in the wind is computationally expensive.
We therefore determine the functional form of $M(t)$ in advance
through the following procedure.

\begin{enumerate}
    \item
    For a given chemical composition, temperature, and density,
    the Saha equations for all ionization stages are solved
    to determine the ionization states.
    These calculations are performed over a range of temperatures
    and densities for each fixed chemical composition, and tables of
    ionization states and level populations are constructed as functions
    of temperature and density (Sec.~\ref{Assec:levpop}).

    \item
    For each temperature and density, the contributions from spectral lines
    associated with all elements and ionization stages are summed using the
    calculated ionization states, thereby determining the functional form
    of $M(t)$ for that temperature and density
    (Sec.~\ref{Assec:Mtcalc}).

    \item
    Following \citet{2021AAS...23711602L},
    the obtained $M(t)$ is fitted using a four-parameter model.
    Tables of the fitting parameters are then constructed
    as functions of temperature and density
    (Sec.~\ref{Assec:Mtfit}).
\end{enumerate}

Using this procedure,
the functional form of $M(t)$ for a given chemical composition
can be obtained directly from the local temperature and density,
without recalculating all line contributions at every point in the wind.
We describe each step in detail in Appendix~\ref{Asec:linemult}.

\subsection{Boundary condition} \label{ssec:boundary}
In addition to the differential equations,
the solution must satisfy several conditions imposed at the inner boundary,
the singular point, and the outer boundary.
Among these, the conditions at the singular point require special treatment and are described below.

The effective opacity $\kappa_{\rm eff}$ in the equation of motion
depends on the velocity gradient through the line force multiplier.
As a result, Eq.~\ref{eq:Euler} cannot be transformed into
an explicit form for $dv/dr$,
and the number of real solutions changes from 0 to 2
depending on the local values of
$r$, $v$, $T$, $\rho$, and $L$.
Because of this property,
a physically continuous wind solution extending from the inner
boundary to the outer boundary must satisfy two conditions
simultaneously at the same radius.
One is the singularity condition,
and the other is the regularity condition
(e.g., \citealt{1975ApJ...195..157C}; see also Appendix~\ref{Asec:itemet}).

The singularity condition corresponds to the point where
the real solutions of the velocity gradient become degenerate,
thereby ensuring the continuity of the wind solution.
The regularity condition guarantees that the velocity gradient
remains continuously defined throughout the wind.
These conditions are written as
\begin{equation}
    \frac{\partial F}{\partial v'}(r_{\rm s})
    =
    0, \label{eq:cond_sing1}
\end{equation}
\begin{equation}
    \frac{dF}{dr}(r_{\rm s})
    =
    0, \label{eq:cond_sing2}
\end{equation}where
\begin{equation}
    F
    \equiv
    v\frac{dv}{dr}
    +
    \frac{1}{\rho}\frac{dp}{dr}
    +
    \frac{GM}{r^2}
    -
    \frac{\kappa_{\rm eff}L}{4\pi r^2 c},
\end{equation}and
\begin{equation}
    v'
    \equiv
    \frac{dv}{dr}.
\end{equation}Here,
$r_{\rm s}$ denotes the radius of the singular point.

The inner boundary is placed at the surface of the degenerate core.
Assuming a zero-temperature and non-rotating degenerate core,
the inner boundary radius is uniquely determined by the degenerate core mass as
\begin{equation}
    r_{\rm in}
    =
    R_{\rm WD}(M_{\rm WD}). \label{eq:cond_in1}
\end{equation}Here, $R_{\rm WD}(M_{\rm WD})$ denotes the mass-radius relation for a cold ONe WD \citep{2019ApJ...887...39K}.
As an inner boundary condition,
we assume that the luminosity at the innermost radius
is supplied by carbon shell burning near the degenerate core surface.
The luminosity is therefore written as
\begin{equation}
    L(r_{\rm in})
    =
    \int_{r_{\rm in}}^{r_{\rm ph}}
    4\pi r^2 \rho \varepsilon_{\rm CC} dr,
\end{equation}where $\varepsilon_{\rm CC}$ is the nuclear energy generation rate
of carbon burning.
Following \cite{2013sse..book.....K},
it is given by
\begin{equation}
\begin{split}
  \varepsilon_{\rm CC}
  &\approx
  \frac{
  5.49\times10^{43}
  \ {\rm erg\,s^{-1}\,g^{-1}}
  \ f_{\rm CC}
  \rho_0
  X_{\rm C}^2
  T_9^{-\frac{3}{2}}
  T_{9\alpha}^{\frac{5}{6}}
  }{
  \exp\left(
  -0.01T_{9\alpha}^{4}
  \right)
  +
  5.56\times10^{-3}
  \exp\left(
  1.685T_{9\alpha}^{\frac{2}{3}}
  \right)
  }
  \\
  &\quad \times
  \exp\left(
  -\frac{84.165}{T_{9\alpha}^{\frac{1}{3}}}
  \right).
\end{split}
\end{equation}Here,
$T_9 = T/(10^9\,{\rm K})$,
$\rho_0 = \rho/(1\,{\rm g\,cm^{-3}})$,
and
\begin{equation}
    T_{9\alpha}
    \equiv
    \frac{T_9}{1.0+0.067T_9}.
\end{equation}For simplicity,
we adopt $f_{\rm CC}=0.5$.
Furthermore, assuming that the nuclear energy generation is strongly concentrated
near the degenerate core surface,
the inner boundary condition can be approximated
using local quantities as
\begin{equation}
    L(r_{\rm in})
    \approx
    4\pi r_{\rm in}^3
    \rho
    \varepsilon_{\rm CC}. \label{eq:cond_in2}
\end{equation}

The outer boundary condition is imposed at an effective photosphere defined by the effective opacity $\kappa_{\rm eff}$.
At this boundary, we require that the luminosity and temperature satisfy the Stefan--Boltzmann law.
Assuming that the wind profile smoothly extends outward,
we define the effective photosphere as
\begin{equation}
    \tau_{\rm eff}
    =
    \kappa_{\rm eff}\rho r_{\rm out}
    =
    \frac{8}{3}, \label{eq:cond_out1}
\end{equation}following the approach of \citet{1994ApJ...437..802K}.
At this location,
the Stefan--Boltzmann law,
\begin{equation}
    L
    =
    4\pi r_{\rm out}^2 \sigma T^4, \label{eq:cond_out2}
\end{equation}is imposed as the outer boundary condition.

\section{Numerical Solver and Parameter Survey Strategy} \label{sec:calmeth}
\subsection{Numerical Solver}
Once the chemical composition is fixed,
the unknown parameters in this system are
$M_{\rm WD}$, $\Lambda_{\rm tot}$, and $\dot{M}$.
In addition, the differential equations
(Eqs.~\ref{eq:Euler} and \ref{eq:effdiffusion})
lead to two integration constants.
Moreover, the locations of the three boundaries must be determined self-consistently,
introducing three additional unknowns.
Consequently, the problem contains eight unknown quantities in total.

The boundary conditions imposed at the three boundaries provide six constraints.
Therefore, two additional physical quantities must be specified to obtain a unique wind solution.

Within this formulation, the wind solution is fully determined by specifying the chemical composition and two additional parameters.
These two quantities are naturally identified with the degenerate core mass $M_{\rm WD}$ and the wind region mass $\Delta M$, which reflect the initial conditions of the binary system and its subsequent secular evolution.
The definition of $\Delta M$ is given by
\begin{equation}
    \Delta M \equiv \int_{r_{\rm in}}^{r_{\rm ph}}4\pi r^2\rho\,dr.
\end{equation}

In practice, however, these quantities are not used as direct control parameters in the numerical calculations.
Instead, we adopt $\Lambda_{\rm tot}$ in place of $\Delta M$ as the free parameter.
This is because, if $\Delta M$ were used as a control parameter, the above integral condition would need to be imposed as an additional constraint, introducing an extra integral restriction into the boundary-value problem.
As a result, the matrix inversion procedure in the relaxation method described below would become unnecessarily complicated, which is undesirable from a numerical standpoint.

On the other hand, when interpreting and discussing the results, we use $(M_{\rm WD}, \Delta M)$ as the physically meaningful parameters characterizing the system.

Two standard approaches for solving boundary-value problems are the shooting method and the relaxation method.
The relaxation method may be regarded as a multidimensional extension of the Newton--Raphson method, in which a trial solution is iteratively improved until convergence is achieved.
A major advantage of this method is that, once a converged solution has been obtained for a given parameter set, neighboring solutions can be computed efficiently by continuously varying the parameters.
This feature is particularly important in the present study, whose main goal is to systematically explore the solution space in the $M_{\rm WD}$–$\Delta M$ plane for each chemical composition and to identify the parameter combinations that reproduce the observed properties of WD J005311.
For this reason, we adopt the relaxation method in this work.
Details of the numerical implementation are given in Appendix~\ref{Asec:itemet}.

\subsection{Parameter Survey Strategy}
We survey solutions for a given chemical composition by varying $M_{\rm WD}$ and $\Lambda_{\rm tot}$.
First, we obtain a converged solution for a reference set of parameters, $M_{\rm WD}^0$ and $\Lambda_{\rm tot}^0$.
This solution is then used as the initial guess for nearby parameter values, $M_{\rm WD}^0 + \delta M_{\rm WD}$ and $\Lambda_{\rm tot}^0 + \delta \Lambda_{\rm tot}$,
and the solution is recomputed until convergence is achieved.
This procedure is repeated to systematically obtain all solutions in the region of parameter space where converged solutions exist.

The radiative acceleration due to continuum radiation strongly depends on the opacity, which is primarily determined by the chemical composition, in particular the metallicity.
In addition, the line-force multiplier also depends on the metallicity, since heavy elements provide a large number of bound–bound transitions that contribute significantly to the line driving.
Therefore, the chemical composition, especially the metallicity, is a key parameter in determining the radiative acceleration.

We explore the parameter space of $M_{\rm WD}$ and $\Lambda_{\rm tot}$ for two representative chemical compositions.
Table~\ref{tb:compmodels} summarizes the two composition models.
\begin{deluxetable}{ccccc}
\tablecaption{Composition Models (Mass fraction)
\label{tb:compmodels}}
\tablehead{
\colhead{Model name} & \colhead{Carbon} & \colhead{Oxygen} & \colhead{Neon} & \colhead{Metallicity except for C,O,Ne\ ($Z$)}
}
\startdata
CO-Solar & $0.200 - 0.5Z$ & $0.700 - 0.5Z$ & $0.100$ & $1.43\times10^{-2}$ \\
Def-Bound & $0.415$ & $0.472$ & $5.24\times10^{-2}$ & $6.06\times10^{-2}$
\enddata
\tablecomments{
For the CO-Solar model, $Z$ denotes the total solar metallicity, including C, O, and Ne.
The abundances listed for C and O represent the values before the solar-metallicity component is added.
Consequently, the final C, O, and Ne mass fractions are larger than the tabulated values because the solar-metallicity component also contains these elements.
}
\end{deluxetable}

The ``CO-Solar'' model represents a gas composed predominantly of carbon, oxygen, and neon with solar metallicity.
The mass fractions of carbon, oxygen and neon are taken from \cite{2019Natur.569..684G}.
The uncertainty in the metallicity remains large (e.g., \citealt{2023ApJ...944..120L}).
Therefore, we adopt solar metallicity as a representative value for the contribution of heavy elements.
The total metallicity excluding carbon, oxygen, and neon is adopted from the proto-solar composition of \cite{2009ARA&A..47..481A},
while the relative abundances of the individual heavy elements are taken from \cite{1989GeCoA..53..197A}.

If WD J005311 is the remnant of a deflagration explosion of a CO WD,
its envelope may consist of a mixture of fallback ejecta and material originating from the bound remnant.
Since the detailed degree of mixing is uncertain,
we adopt the limiting case in which the wind region is entirely composed of bound-remnant material.
This composition is referred to as the ``Def-Bound'' model.
The adopted elemental abundances are taken from the bound-remnant composition of the ``N1def'' model of \cite{2014MNRAS.438.1762F}.
The detailed elemental composition is summarized in Table~\ref{tb:detailcomp}.
\begin{deluxetable}{cccc}
\tablecaption{Mass fractions of individual elements adopted in each model \label{tb:detailcomp}}
\tablehead{
 & \colhead{CO-Solar} & \colhead{Def-Bound} 
}
\startdata
H  & $0.00$ & $0.00$ \\
He & $0.00$ & $2.16\times10^{-5}$ \\
Li & $7.53\times10^{-9}$ & $0.00$ \\
Be & $1.25\times10^{-10}$ & $0.00$ \\
B  & $4.37\times10^{-9}$ & $0.00$ \\
C  & $0.195$ & $0.415$ \\
N  & $8.35\times10^{-4}$ & $2.42\times10^{-4}$ \\
O  & $0.700$ & $0.472$ \\
F  & $3.05\times10^{-7}$ & $9.41\times10^{-9}$ \\
Ne & $0.101$ & $5.24\times10^{-2}$ \\
Na & $2.52\times10^{-5}$ & $2.13\times10^{-3}$ \\
Mg & $4.97\times10^{-4}$ & $1.20\times10^{-2}$ \\
Al & $4.37\times10^{-5}$ & $1.85\times10^{-3}$ \\
Si & $5.35\times10^{-4}$ & $9.73\times10^{-3}$ \\
P  & $6.14\times10^{-6}$ & $1.15\times10^{-4}$ \\
S  & $3.15\times10^{-4}$ & $2.46\times10^{-3}$ \\
Cl & $2.55\times10^{-6}$ & $1.14\times10^{-5}$ \\
Ar & $6.99\times10^{-5}$ & $3.66\times10^{-4}$ \\
K  & $2.62\times10^{-6}$ & $3.13\times10^{-6}$ \\
Ca & $4.72\times10^{-5}$ & $2.50\times10^{-4}$ \\
Sc & $2.93\times10^{-8}$ & $4.41\times10^{-9}$ \\
Ti & $2.19\times10^{-6}$ & $6.95\times10^{-6}$ \\
V  & $2.84\times10^{-7}$ & $2.27\times10^{-6}$ \\
Cr & $1.34\times10^{-5}$ & $2.38\times10^{-4}$ \\
Mn & $1.00\times10^{-5}$ & $4.36\times10^{-4}$ \\
Fe & $9.58\times10^{-4}$ & $2.83\times10^{-2}$ \\
Co & $2.53\times10^{-6}$ & $1.20\times10^{-5}$ \\
Ni & $5.53\times10^{-5}$ & $2.63\times10^{-3}$ \\
Cu & $6.33\times10^{-7}$ & $2.01\times10^{-8}$ \\
Zn & $1.57\times10^{-6}$ & $1.78\times10^{-7}$ \\
\enddata
\end{deluxetable}

\section{Result} \label{sec:result}
\subsection{Wind Profiles}
\begin{figure}[t]
\centering
\includegraphics[width=0.4\textwidth]{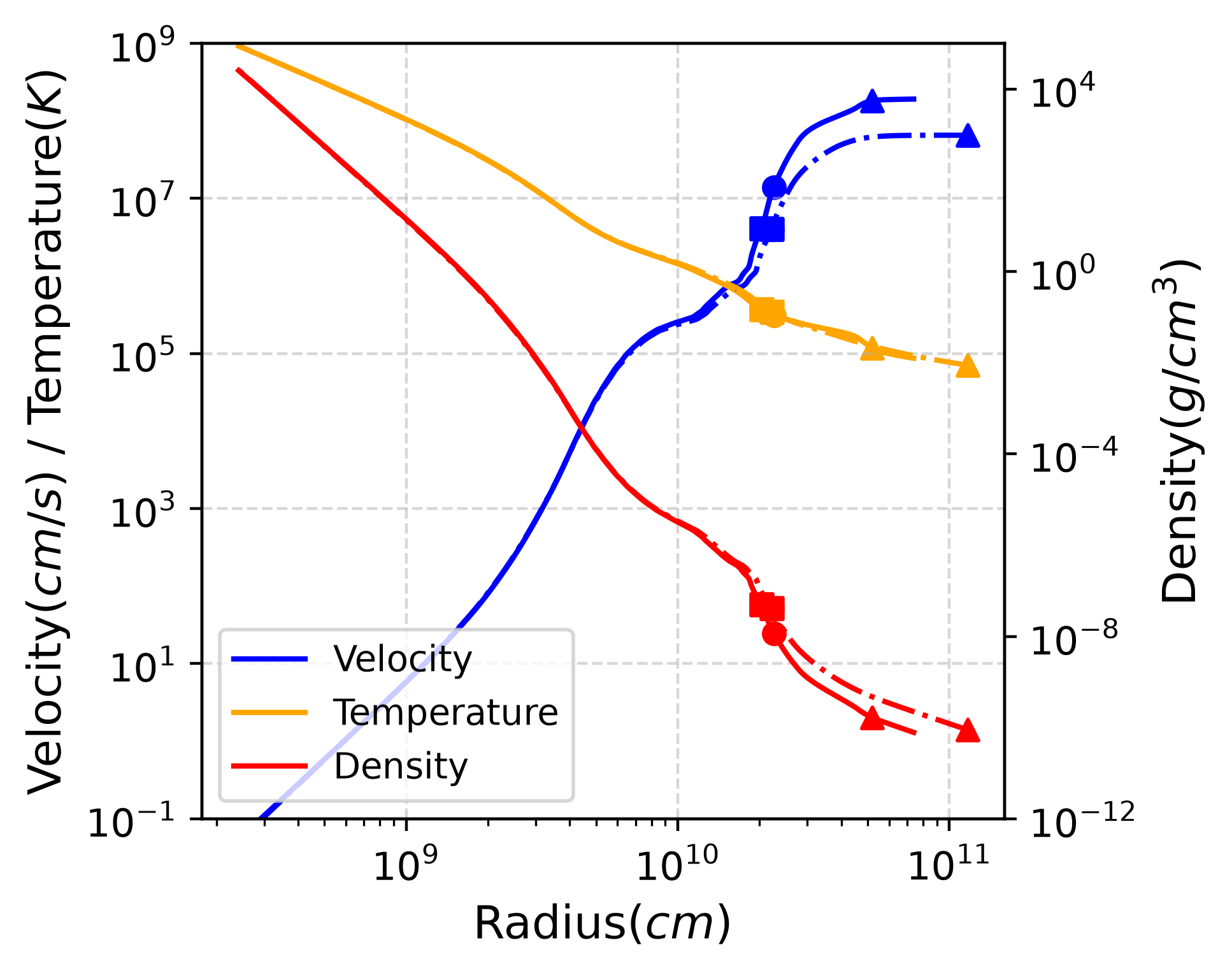}
\caption{
Wind profiles for the CO-Solar model with
$M_{\rm WD}=1.30\,M_{\odot}$ and
$\Delta M=1.00\times10^{-3}\,M_{\odot}$.
The blue, orange, and red lines show the velocity, temperature, and density profiles, respectively.
To illustrate the contribution of the line force, the corresponding continuum-driven wind solution with the same
$M_{\rm WD}$ and $\Delta M$ is shown by dash-dotted lines.
Square symbols denote the transonic points,
circle symbols denote the singular points,
and triangle symbols denote the photosphere defined using the continuum Rosseland mean opacity.
\label{fg:ex_wind}}
\end{figure}

Before presenting the overall properties of the wind solutions, we first examine a representative example.
Fig.~\ref{fg:ex_wind} shows the velocity, density, and temperature profiles of a representative wind solution.

This model adopts the CO-Solar composition with
$M_{\rm WD}=1.30\,M_{\odot}$ and
$\Delta M=1.00\times10^{-3}\,M_{\odot}$.
Square, circle, and triangle symbols denote the transonic point, the singular point, and the photosphere defined using the continuum Rosseland mean opacity, respectively.
The photosphere is defined by the condition
\begin{equation}
    \tau_{\rm R} \equiv \int_{r_{\rm ph}}^{\infty} \kappa_{\rm R} \rho \, dr = \frac{2 }{3 }.
\end{equation}Outside the photosphere, the wind is optically thin with respect to the continuum Rosseland mean opacity, while it remains optically thick when the effective opacity $\kappa_{\rm eff}$, including the line force, is adopted.

To illustrate the contribution of the line force, we also plot the corresponding continuum-driven wind solution, in which only the continuum radiation force is included, as dash-dotted lines.
For comparison, we select the continuum-driven solution with the same $M_{\rm WD}$ and $\Delta M$.
The formulation of the continuum-driven solutions is described in Appendix~\ref{Asec:contonly}.

This figure shows that the terminal wind velocity is higher in the line-driven solution than in the continuum-driven solution.
The higher wind velocity leads to a lower density, and consequently, the photosphere shifts to a deeper (more inward) location.

\begin{figure}[t]
\centering
\includegraphics[width=0.4\textwidth]{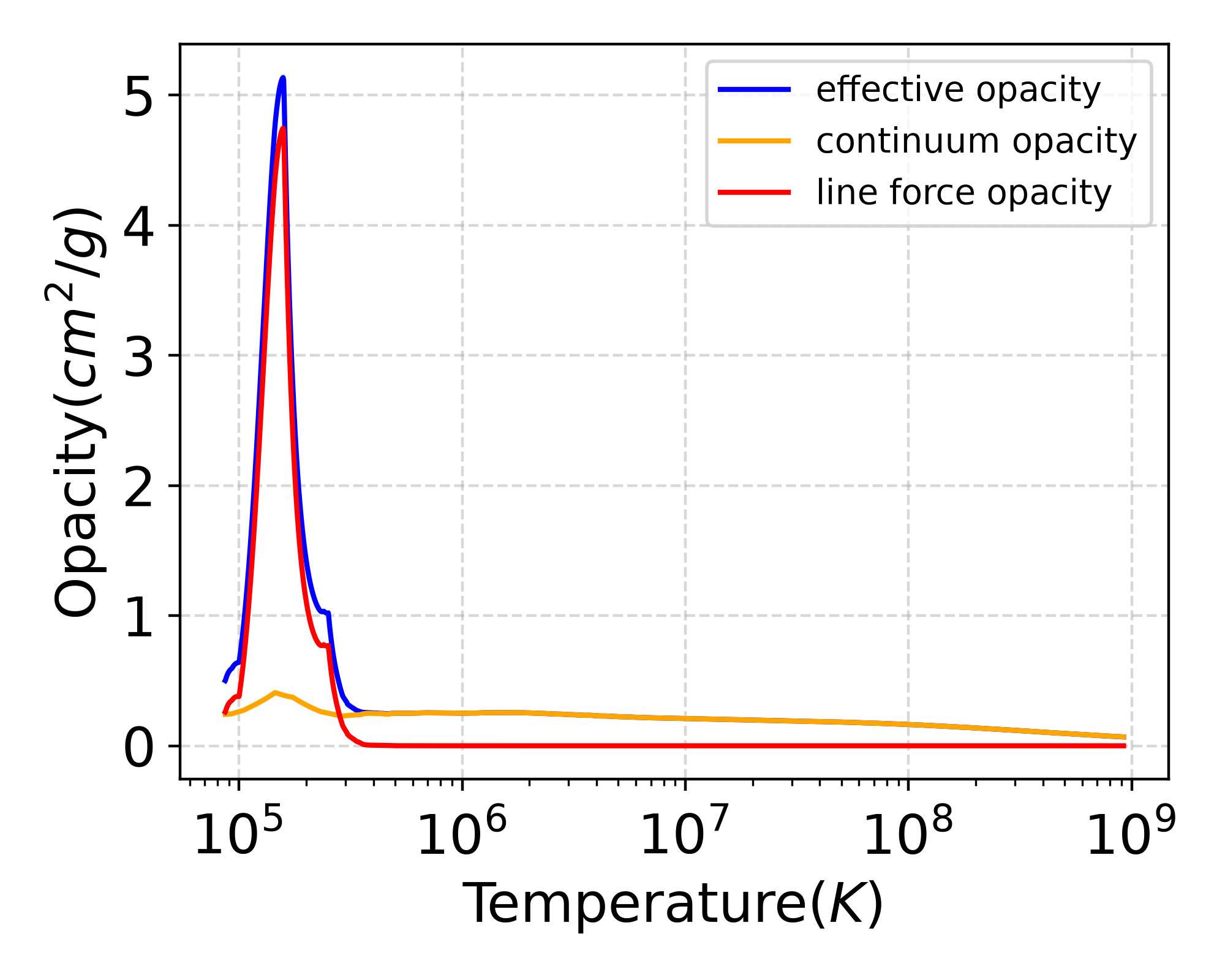}
\caption{
Temperature-opacity relation for the solution shown in Fig.~\ref{fg:ex_wind}.
The blue, orange, and red lines represent the effective opacity, the continuum opacity, and the line opacity, respectively.
\label{fg:ex_opacity}}
\end{figure}

Fig.~\ref{fg:ex_opacity} shows the temperature-opacity relation for this solution.
In the inner region, the continuum opacity is larger than the line opacity.
However, outside the photosphere, the line opacity exceeds the continuum opacity.
Therefore, the region outside the photosphere can be regarded as a line force dominated region.

Having examined the properties of a representative wind solution, we now turn to the overall behavior of the solution sequence for the CO-Solar model, in which the degenerate core mass is fixed while the wind region mass is varied.

For large values of $\Delta M$, the acceleration region is located well inside the photosphere, where the line force is ineffective.
As a result, the wind is accelerated solely by continuum radiation.
In our calculations, the adopted approximation for the line force does not allow the Euler equation to describe solutions with a negative velocity gradient.
Therefore, no physically meaningful line-driven solutions are obtained in this range of $\Delta M$, and the continuum-driven solutions are regarded as the physically relevant solutions.

Conversely, for small values of $\Delta M$, the photospheric radius decreases, bringing the acceleration region closer to the photosphere.
In this case, the region above the photosphere has a large velocity gradient, making the line force more effective.
Consequently, physically meaningful line-driven solutions exist in this range.
On the other hand, in the continuum-driven scheme, the photosphere is located too deep for the wind to be accelerated to the escape velocity.
As a result, no continuum-driven solutions are obtained in this range of $\Delta M$.

In summary, the continuum-driven and line-driven solutions play
complementary roles over the range from large to small values of $\Delta M$.
The optically thick wind phase consists of these two types of wind solutions.
Although they play complementary roles, the two types of solutions are not
mutually exclusive: there are ranges of $\Delta M$ where both
continuum-driven and line-driven wind solutions exist.
For example, Fig.~\ref{fg:ex_wind} compares these two types of solutions
in such a range.

\begin{figure*}[t]
\centering
\includegraphics[width=1.0\textwidth]{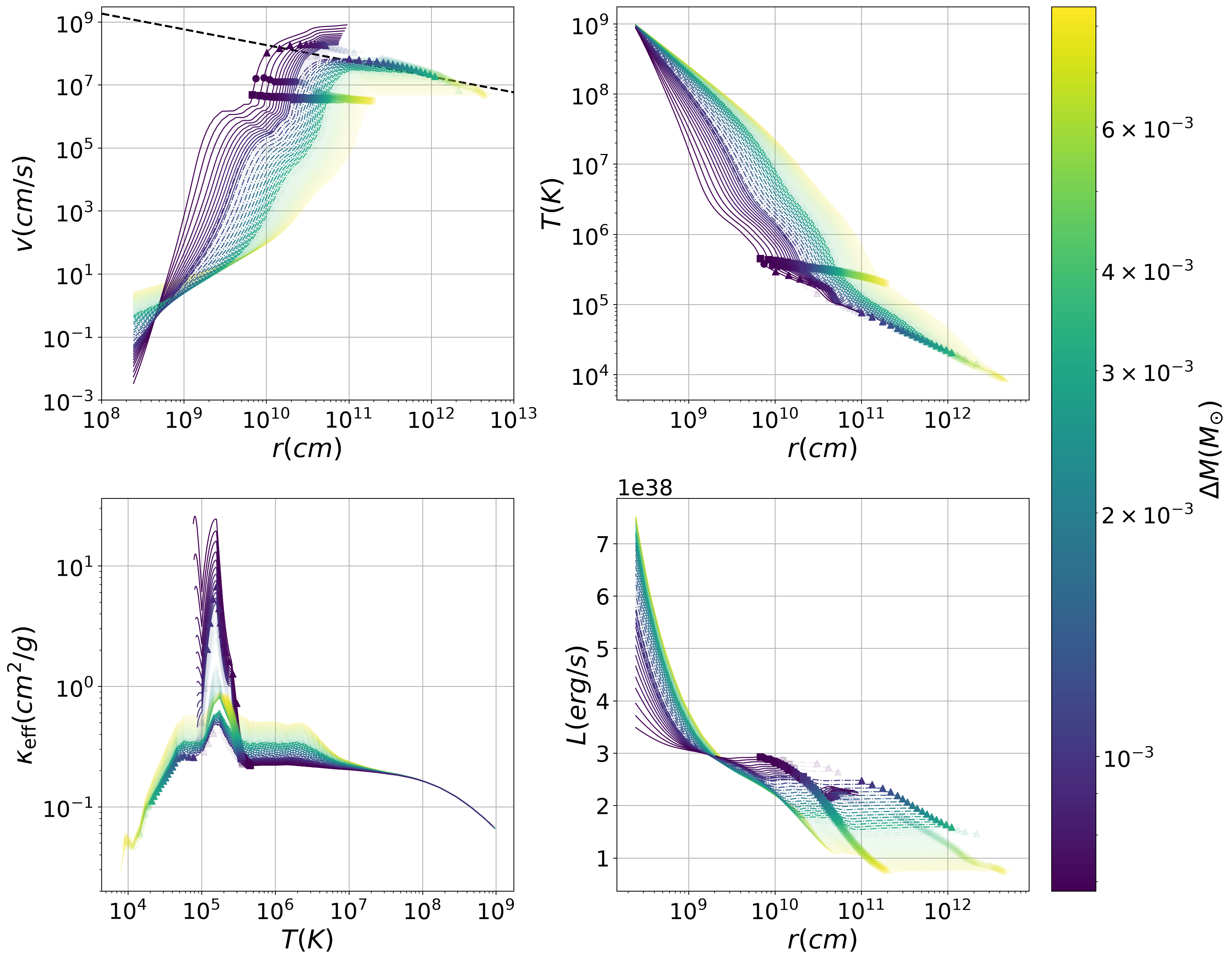}
\caption{
Solution sequence for the CO-Solar model with $M_{\rm WD}=1.3\,M_{\odot}$ and different wind region masses, $\Delta M$.
The upper-left panel shows the velocity profiles, the upper-right panel the temperature profiles, the lower-left panel the effective opacity as a function of temperature, and the lower-right panel the luminosity profiles.
The color of each curve represents $\Delta M$, as indicated by the color bar.
To illustrate the continuity of the sequence, continuum-driven wind solutions are shown together with the line-driven wind solutions; solid lines represent line-driven wind solutions, whereas dash-dotted lines represent continuum-driven wind solutions.
Solutions that converged numerically but either entered a parameter regime where the Euler equation has no physically valid solution (see Appendix~\ref{Asec:itemet}) or failed to reach the local escape velocity at the outer boundary are shown with semi-transparent lines.
Square, circle, and triangle symbols indicate the transonic points, singular points, and photospheres defined using the continuum Rosseland mean opacity, respectively.
\label{fg:COS_DeltaMseq}}
\end{figure*}

Based on the above discussion, Fig.~\ref{fg:COS_DeltaMseq} shows the solution sequence for a fixed degenerate core mass, $M_{\rm WD}=1.3\,M_{\odot}$, 
constructed by combining the continuum-driven and line-driven solutions over the range of $\Delta M$ for which physically meaningful solutions can be obtained.
We show the velocity, temperature, and luminosity profiles, together with the effective opacity as a function of temperature, for each solution.

The range of $\Delta M$ represents the parameter space in which converged wind solutions are obtained.
The upper limit corresponds to the maximum wind region mass for which radiation pressure can still accelerate the wind to supersonic velocities.
In contrast, the lower limit corresponds to the minimum wind region mass for which the photosphere remains outside the transonic point.
Note that, in general, the boundary-value problem becomes increasingly difficult to solve as $\Delta M$ approaches either limit.
Therefore, the limits presented here can be regarded as approximate physical boundaries of the solution space, but may not exactly coincide with the physical boundaries.

When $\Delta M$ is small, the positions of the sonic point and the critical point shift inward, 
making the flow more susceptible to acceleration and enhancing the effect of the line force.
Consequently, the smaller the $\Delta M$, the higher the terminal wind velocity. 
If the sequence of $\Delta M$ values is viewed as representing temporal evolution driven by mass-loss,
it follows that the terminal wind velocity reaches its maximum during the final stages of evolution.

\begin{figure*}[t]
\centering
\includegraphics[width=0.9\textwidth]{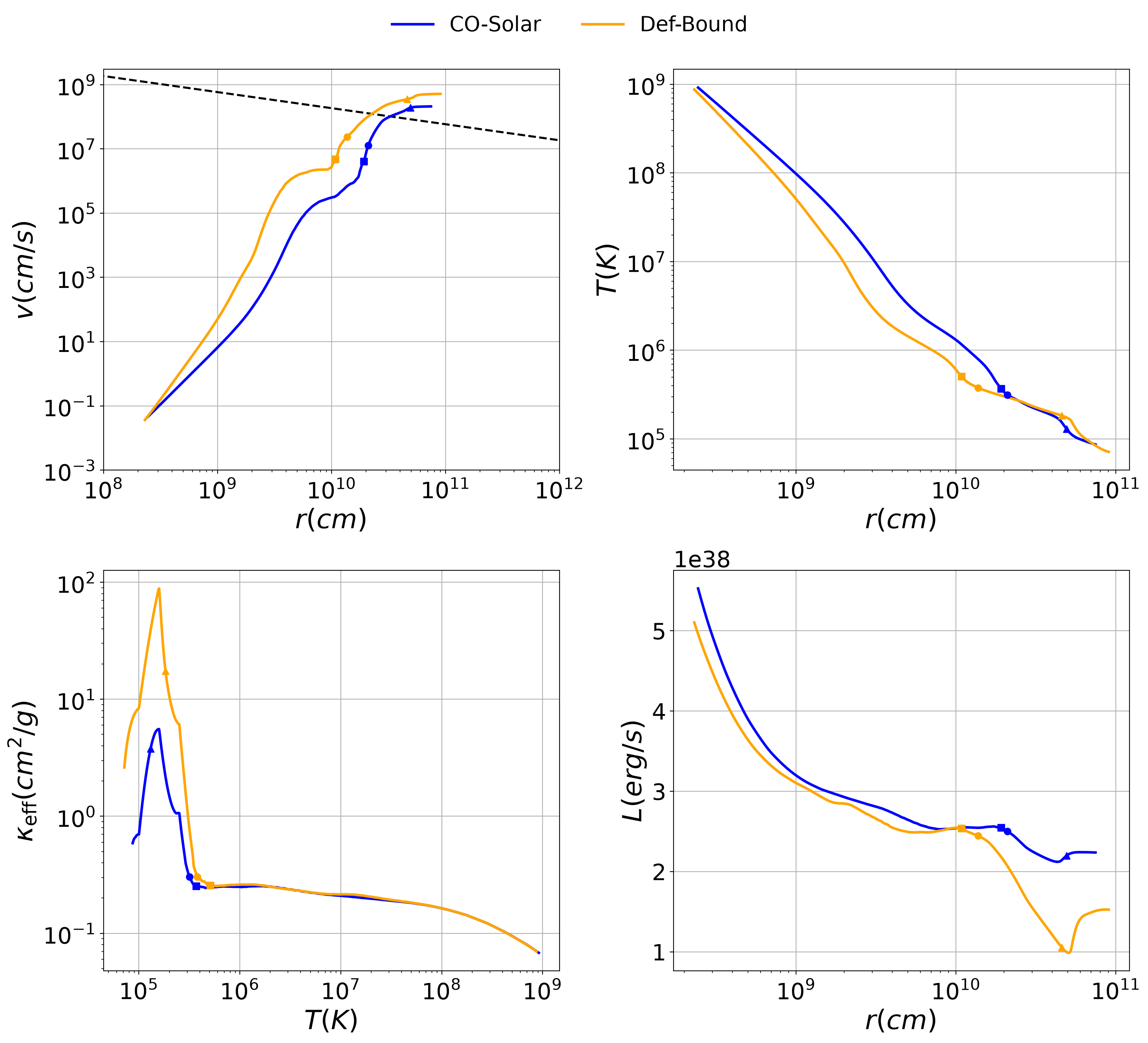}
\caption{
Comparison of wind solutions for different chemical composition models.
The two solutions have the same $M_{\rm WD}=1.3\,M_{\odot}$ and $\Lambda_{\rm tot}=2.5\times10^{38}\ {\rm erg\,s^{-1}}$, but different chemical compositions.
The blue and orange curves represent the CO-Solar and Def-Bound models, respectively.
Each panel shows the same quantities as in Fig.~\ref{fg:COS_DeltaMseq}.
\label{fg:compcomp}}
\end{figure*}

Next, we turn to the dependence of the wind solutions on the chemical composition.
Fig.~\ref{fg:compcomp} compares representative wind solutions with the same degenerate core mass, $M_{\rm WD}$, and total luminosity, $\Lambda_{\rm tot}$, but different chemical composition models.
We compare solutions with the same $M_{\rm WD}$ and $\Lambda_{\rm tot}$, rather than the same $\Delta M$, because $\Lambda_{\rm tot}$ directly measures the radiative power available to accelerate the wind.
Matching both $M_{\rm WD}$ and $\Lambda_{\rm tot}$ therefore provides a more appropriate basis for evaluating the influence of the chemical composition on the wind acceleration.

Fig.~\ref{fg:compcomp} shows that the terminal wind velocity of the Def-Bound model is larger than that of the CO-Solar model.
This trend indicates that a higher abundance of heavy elements increases both the continuum and line opacities, allowing the radiation field to transfer more momentum to the wind for a given radiative power.
This interpretation is also supported by the larger values of $\kappa_{\rm eff}$ in the low-temperature region and the stronger reduction in luminosity in the outer wind for the Def-Bound model compared with the CO-Solar model.

\subsection{Global Wind Properties}
\begin{figure*}[t]
\centering
\includegraphics[width=0.8\textwidth]{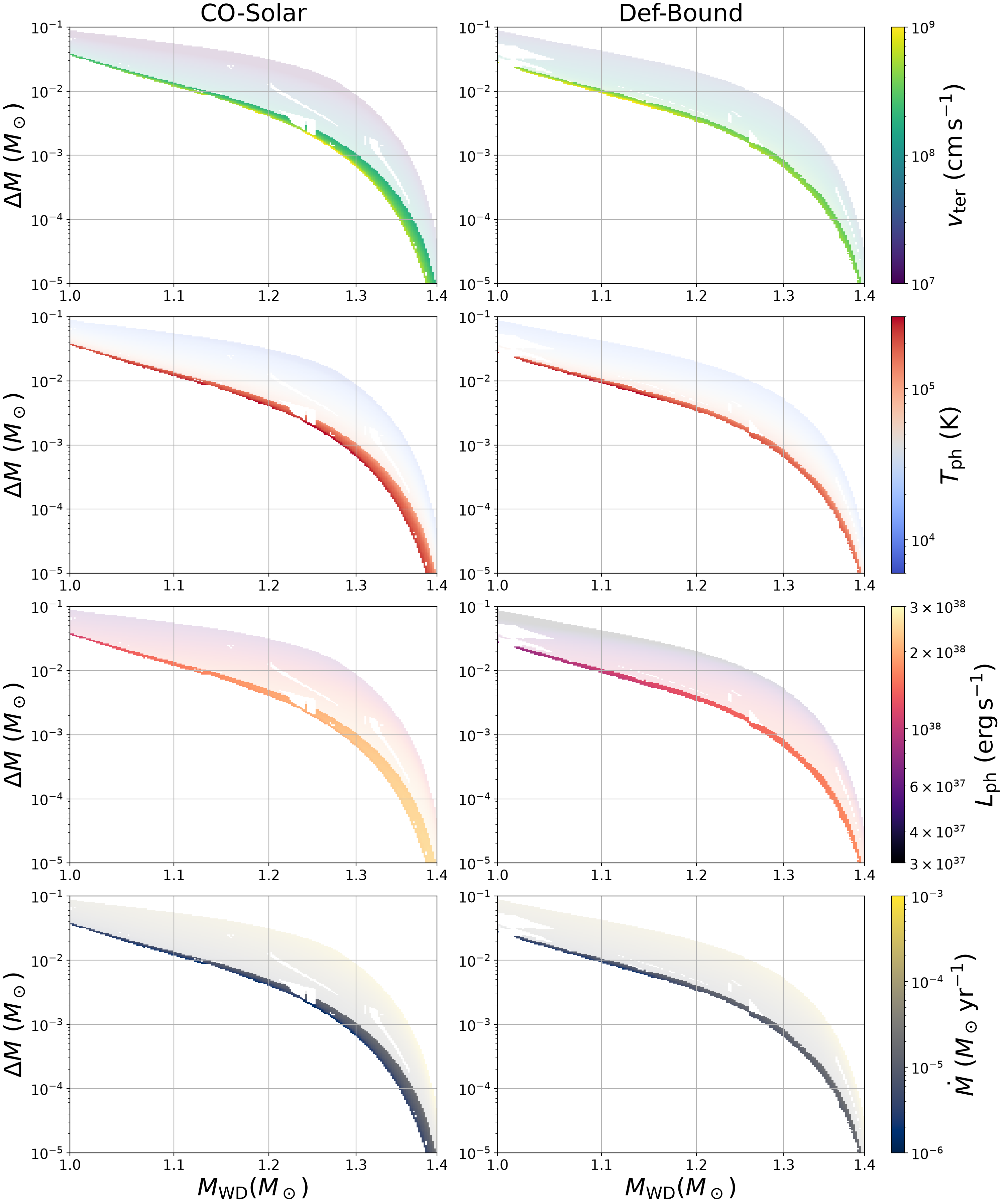}
\caption{
Distributions of some physical quantities for the line-driven wind solutions on the $M_{\rm WD}$--$\Delta M$ plane for the two chemical composition models.
From left to right, the columns correspond to the CO-Solar and Def-Bound models.
The top row shows the terminal wind velocity at the outer boundary.
The second and third rows show the photospheric effective temperature and luminosity, respectively, where the photosphere is defined as the location at which the optical depth measured inward from the outer boundary using the Rosseland mean opacity reaches $2/3$.
The bottom row shows the mass-loss rate.
Regions containing solutions that converged numerically but either entered a parameter regime where the Euler equation has no physically valid solution or failed to reach the local escape velocity at the outer boundary are shown with semi-transparent colors.
\label{fg:vTLMdot_2models_effOTW}}
\end{figure*}

In this subsection, we examine how the global properties of the optically thick wind solutions depend on the model parameters.
Fig.~\ref{fg:vTLMdot_2models_effOTW} shows the distributions of the terminal wind velocity, effective temperature, luminosity, and mass-loss rate on the $M_{\rm WD}$--$\Delta M$ plane for the two chemical composition models.
The terminal wind velocity is evaluated at the outer boundary, 
whereas the effective temperature and the luminosity are evaluated at the
photosphere, which is defined as the location where the optical depth
measured inward from the outer boundary, using the Rosseland mean
opacity, reaches $2/3$.
This choice is motivated by the fact that the effective temperature
inferred from observations corresponds to the photospheric value.

\begin{figure*}[t]
\centering
\includegraphics[width=0.8\textwidth]{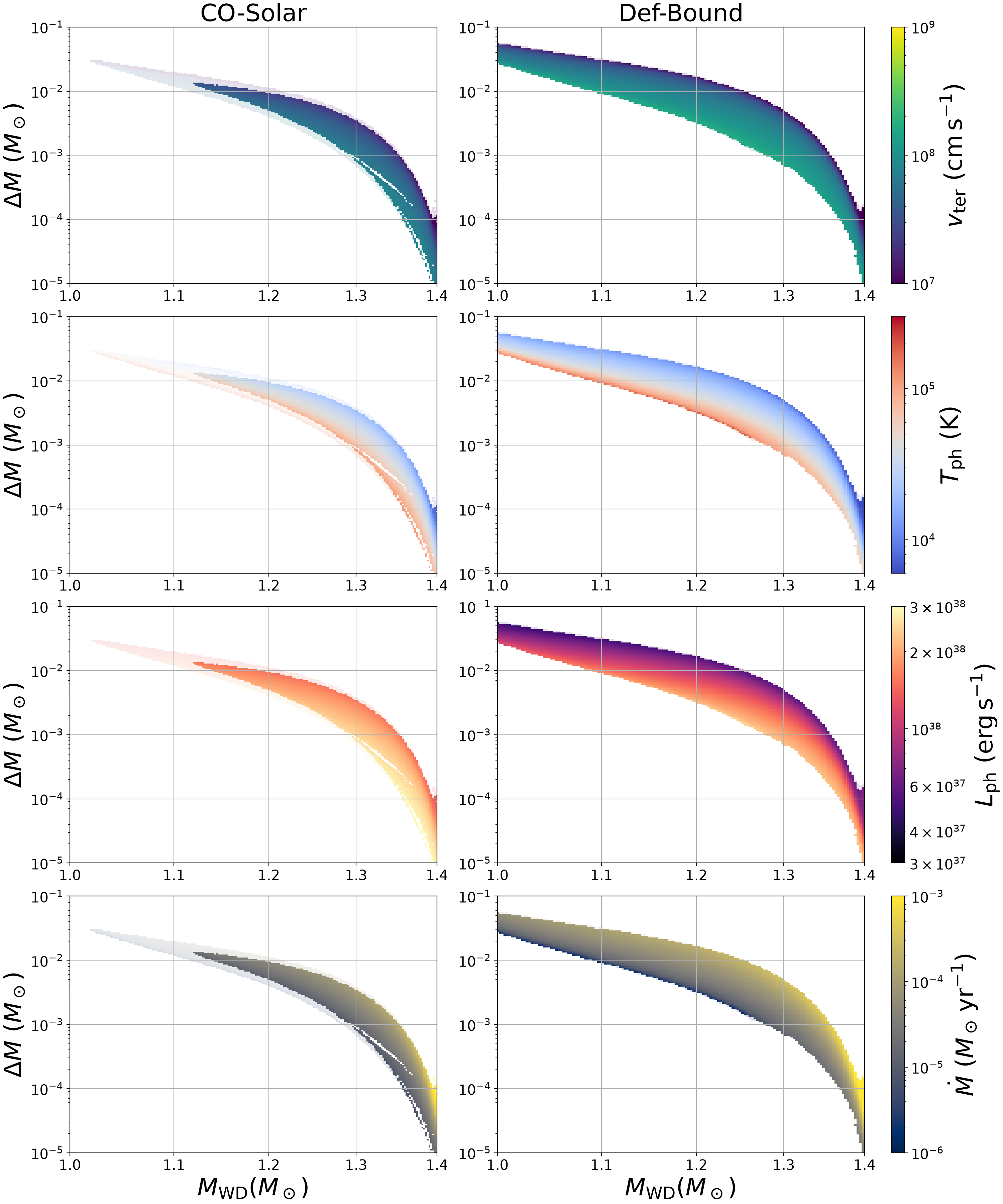}
\caption{
Same as Fig.~\ref{fg:vTLMdot_2models_effOTW}, but for the continuum-driven wind solutions.
In the continuum-driven regime, the photosphere is identified with the outer boundary.
Since the line force is negligible in this regime, the velocity at the photosphere is regarded as the terminal wind velocity.
Solutions that failed to reach the local escape velocity are shown with semi-transparent colors.
\label{fg:vTLMdot_2models_OTW}}
\end{figure*}

Fig.~\ref{fg:vTLMdot_2models_effOTW} shows that the range of $\Delta M$ over which line-driven wind solutions exist is relatively narrow.
The lower limit of $\Delta M$ is determined by the condition that the photosphere remains outside the transonic point.
The upper limit is determined by the limitation of our line-driven wind formulation, which cannot describe solutions with a negative velocity gradient.
As discussed above, continuum-driven wind solutions are regarded as the physically relevant solutions in the parameter range where our line-driven wind formulation is not applicable because the line force is negligible.
For comparison, the corresponding distributions for the continuum-driven wind solutions are shown in Fig.~\ref{fg:vTLMdot_2models_OTW}.
Together, Figs.~\ref{fg:vTLMdot_2models_effOTW} and \ref{fg:vTLMdot_2models_OTW} demonstrate that the continuum-driven and line-driven wind solutions play complementary roles over the parameter space.
The line-driven solutions are limited to a relatively narrow range of $\Delta M$, while the continuum-driven solutions extend to larger $\Delta M$.
The upper boundary of the overall wind solution sequence is determined by the maximum $\Delta M$ for which the continuum-driven wind can still be accelerated beyond the escape velocity.

The upper and lower limits of $\Delta M$ for which line-driven and continuum-driven
wind solutions exist both shift toward smaller values as the degenerate core mass
increases.
This is because a more massive degenerate core has a deeper gravitational
potential, making it increasingly difficult for optically thick winds with larger
wind region masses to be accelerated beyond the escape velocity.

If the wind region mass of the merger remnant exceeds the maximum
$\Delta M$ for its degenerate core mass, the outer envelope can no
longer be maintained as an optically thick wind.
This corresponds to the hydrostatic giant phase discussed in the Introduction.
Conversely, if the wind region mass is smaller than the lower limit of
$\Delta M$, the wind becomes optically thin, leaving the degenerate core nearly
exposed.

An important result is that optically thick wind solutions are found only for
degenerate core masses above approximately $1\,M_\odot$.
Below this threshold, no optically thick wind solutions are obtained for any
wind region mass considered in this study.
This suggests that, regardless of the wind region mass, merger remnants with
lower-mass degenerate cores are unlikely to form the dynamically outflowing
optically thick winds assumed in the present model.
Instead, they are expected to possess extended, hydrostatic envelopes.

For each fixed degenerate core mass, the terminal wind velocity decreases with increasing $\Delta M$.
This trend is consistent with the discussion of Fig.~\ref{fg:COS_DeltaMseq}.
On the other hand, the maximum terminal wind velocity attained for each degenerate core mass increases with the degenerate core mass.
This is because a more massive degenerate core has a smaller radius, resulting in a higher radiative flux at the inner boundary and hence a stronger radiative acceleration.
This trend becomes less clear for the most massive models
($M_{\rm WD}=1.3$--$1.4\,M_\odot$), because numerical limitations
of the present formulation prevent us from fully exploring the solutions
toward lower $\Delta M$, and thus from obtaining solutions at the true
physical lower limit of $\Delta M$.

We next compare the results of the CO-Solar and Def-Bound models.
The Def-Bound model exhibits a wider range of $\Delta M$ over which optically thick wind solutions exist than the CO-Solar model.
The lower limit of $\Delta M$ shifts to smaller values because the higher opacity allows the photosphere to remain outside the transonic point even for less massive wind regions.
At the same time, the upper limit shifts to larger values because the stronger radiative acceleration can support more massive wind regions against gravity.

The Def-Bound model also produces faster winds than the CO-Solar model, particularly for lower degenerate core masses.
This is a consequence of its metal-rich composition, which enhances the line force and makes the radiative acceleration more efficient.
In contrast, the photospheric luminosity is systematically lower in the Def-Bound model.
Because radiation is absorbed more efficiently, a smaller luminosity is sufficient to accelerate the wind to an unbound state.
On the other hand, the effective temperature and the mass-loss rate show no significant differences between the two models.

The most important implication of these results is that merger remnants are capable of launching ultrafast winds with terminal velocities exceeding
$\sim10^9\,{\rm cm\,s^{-1}}$, comparable to that inferred for WD J005311,
without invoking stellar rotation or magnetic fields.
This demonstrates that radiation-driven optically thick winds alone provide a viable mechanism for producing such high-velocity winds over an appropriate range of system parameters.

\subsection{Time Evolution of Merger Remnant} \label{ssec:timeevolution}
Merger remnants in the optically thick wind phase evolve primarily through mass loss from the wind region. We neglect the growth of the degenerate core due to nuclear burning because its timescale is sufficiently longer than the mass-loss timescale considered here. Therefore, the evolution of a merger remnant can be approximated by a sequence of steady-state solutions with progressively decreasing mass in the wind region.

We connect the evolutionary time to the mass of the wind region using
\begin{equation}
    \Delta M\left( t + \Delta t \right)
    =
    \Delta M\left( t \right)
    -
    \dot{M}(\Delta M \left( t \right)) \Delta t,
\end{equation}where $\dot{M}$ is the mass-loss rate obtained from each steady-state solution. By applying this relation successively along the sequence of solutions, we can estimate the time evolution of a merger remnant for a given degenerate core mass.

\begin{figure*}[t]
\centering
\includegraphics[width=0.8\textwidth]{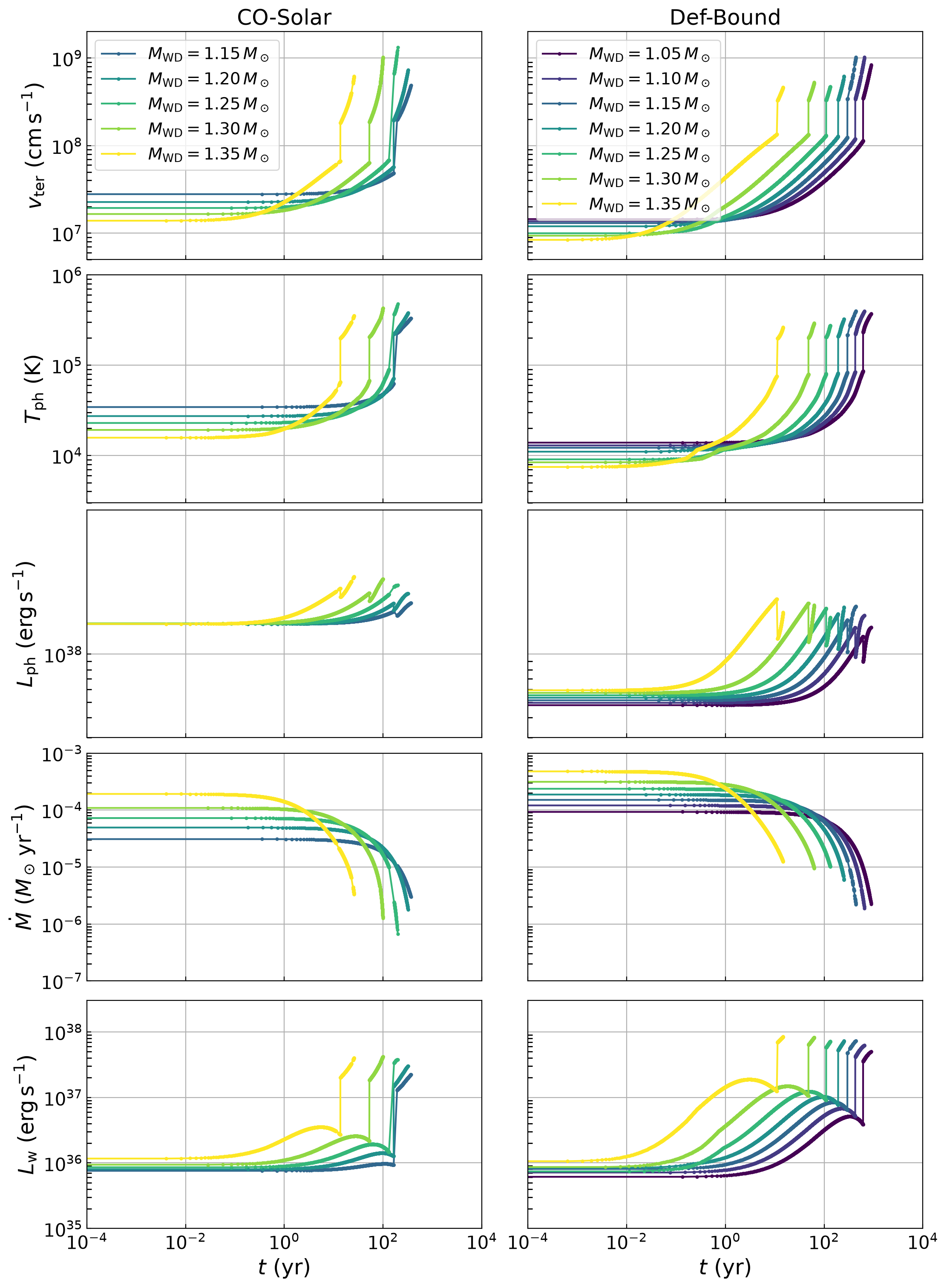}
\caption{Time evolution of some physical quantities for different
degenerate core masses.
The left and right columns show the results for the CO-Solar and
Def-Bound chemical compositions, respectively.
From top to bottom, the panels show the terminal wind velocity,
effective temperature, photospheric luminosity, mass-loss rate, and
wind kinetic luminosity.
Each line with markers represents the evolutionary track of a merger
remnant for a given degenerate core mass.
The tracks are constructed by connecting the continuum-driven and
line-driven wind solutions.
The origin of time is defined as the point corresponding to the
solution with the largest wind-region mass in the continuum-driven phase.
\label{fg:evovTLMdotLw}}
\end{figure*}

Fig.~\ref{fg:evovTLMdotLw} shows the resulting time evolution of some physical quantities for different degenerate core masses.
We define the origin of time as the beginning of the optically thick
wind phase, which corresponds to the boundary between the hydrostatic
giant phase and the optically thick wind phase.
The continuum-driven and line-driven solutions are connected so that
the wind-region mass $\Delta M$ varies continuously. When both solutions
exist for the same $\Delta M$, we preferentially adopt the line-driven
solution. As a result, apparent discontinuities can arise at the
connection points in the evolutionary tracks. These apparent
discontinuities are likely a consequence of the way the two types of
solutions are connected, and the wind is expected to transition
smoothly from the continuum-driven to the line-driven phase.

This figure shows that the duration of the optically thick wind phase decreases with increasing degenerate core mass.
For a degenerate core mass of approximately $1.35\,M_{\odot}$, the lifetime of this phase is only about $10~{\rm yr}$, whereas the lifetime is about $1000~{\rm yr}$ for a degenerate core mass of approximately $1.05\,M_{\odot}$.
Although the two chemical composition models have different wind properties, the overall evolutionary behavior is similar between the two models.
In particular, the dependence of the lifetime of the optically thick wind phase on the degenerate core mass is qualitatively the same for both composition models.

As the mass in the wind region decreases, the photospheric radius
decreases and the effective temperature increases, while the
photospheric luminosity remains nearly constant.
Therefore, the merger remnant moves approximately horizontally to the
left on the HR diagram during the optically thick wind phase.
Although the mass-loss rate decreases with time as the photospheric
radius and density decrease, the wind kinetic luminosity,
$L_{\rm w}\equiv (1/2)\dot{M}v_{\rm ter}^2$, increases as the terminal
wind velocity $v_{\rm ter}$ increases, particularly during the
line-driven wind phase.

\begin{figure*}[t]
\centering
\includegraphics[width=0.8\textwidth]{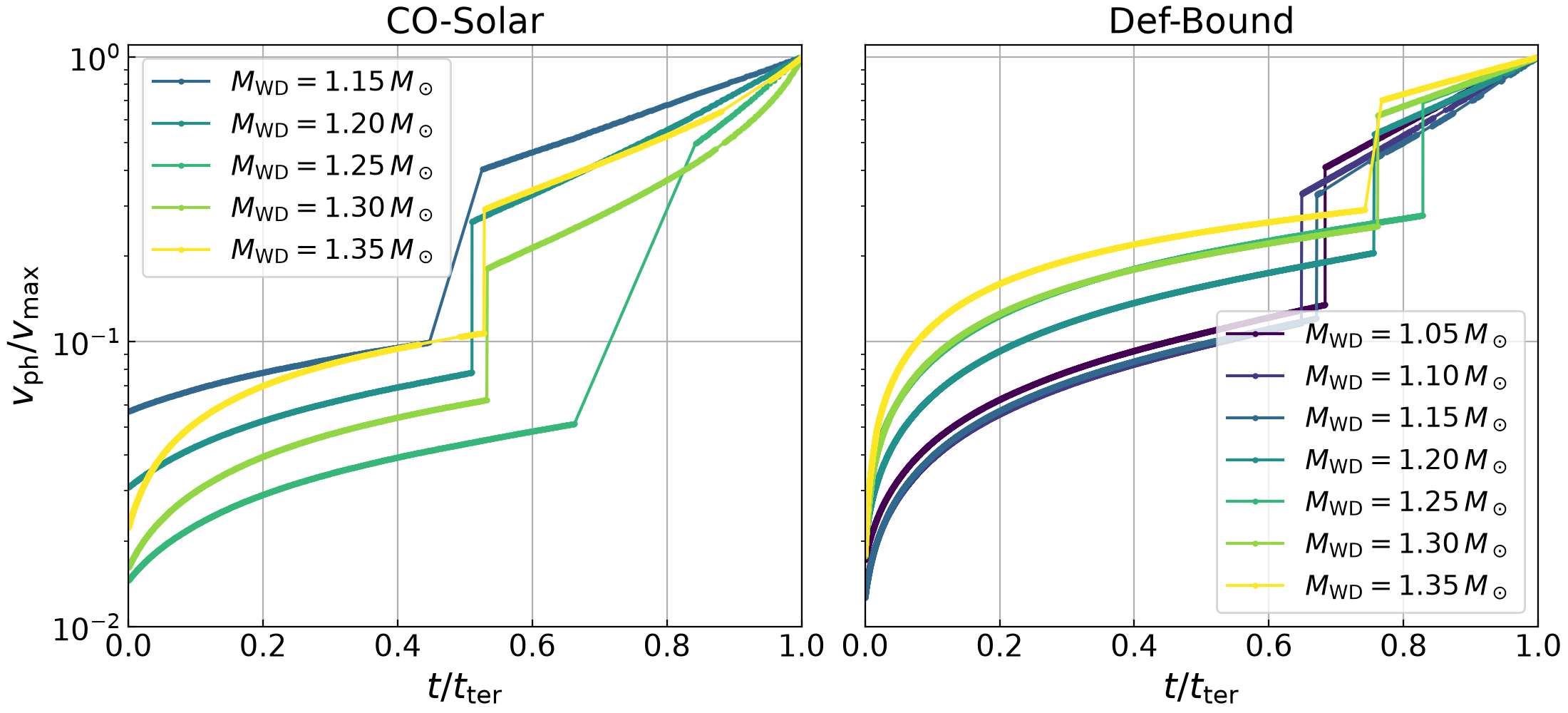}
\caption{Time evolution of the terminal wind velocity normalized by its maximum value for different degenerate core masses.
The left and right panels show the results for the CO-Solar and Def-Bound chemical compositions, respectively.
Each line with markers represents the evolutionary track of a merger remnant with a given degenerate core mass.
The horizontal axis shows the evolutionary time normalized by the lifetime of the optically thick wind phase, while the vertical axis shows the terminal wind velocity normalized by its maximum value.
The results are constructed by connecting the continuum-driven and line-driven wind solutions.
The origin of time is defined as the point corresponding to the solution with the largest wind-region mass in the continuum-driven phase.
\label{fg:evovnondim}}
\end{figure*}

Since the mass-loss rate decreases with time, a given amount of mass is lost over a longer timescale at later stages.
Fig.~\ref{fg:evovnondim} shows the time evolution of the terminal wind velocity normalized by its maximum value for different degenerate core masses.
In the CO-Solar model case, the transition occurs at approximately $50\%$ of the total lifetime for all degenerate core masses.
In the Def-Bound model case, the continuum-driven phase is relatively longer because the higher opacity allows a larger mass to be sustained in the wind region.
Nevertheless, the line-driven phase still accounts for a few tens of percent of the total lifetime.

These results indicate that the line-driven phase is important from the perspective of evolutionary timescales.
Although the region where line-driven wind solutions exist in the $M_{\rm WD}-\Delta M$ plane is relatively narrow, merger remnants can spend a substantial fraction of their lifetime in this phase.
Therefore, the line-driven phase should not be regarded as negligible based solely on the extent of its solution region in parameter space.

\section{Discussion} \label{sec:discussion}
\subsection{Constraints on WD J005311} \label{ssec:constWDJ005311}
\begin{figure*}[t]
\centering
\includegraphics[width=0.8\textwidth]{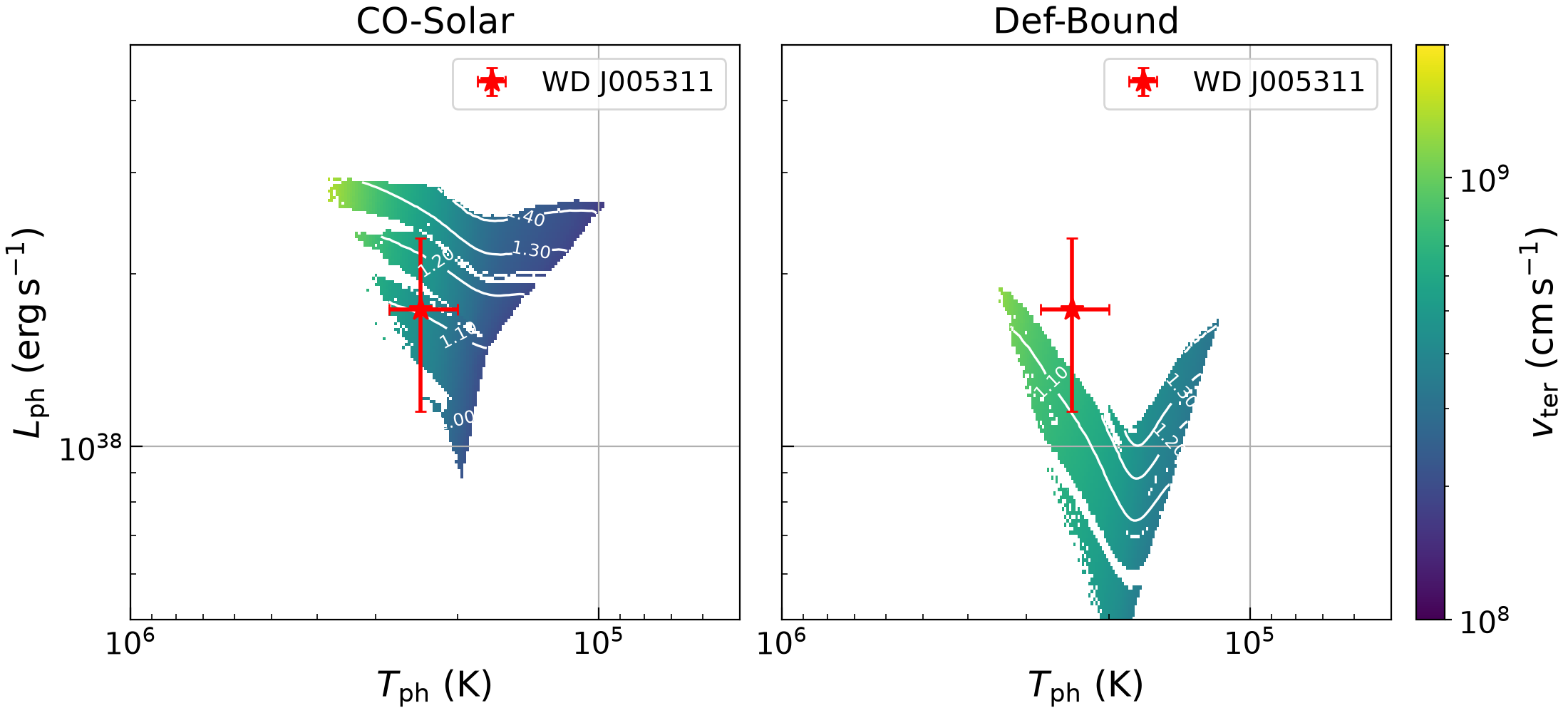}
\caption{
HR diagrams comparing the numerical solutions with the observational constraints for WD J005311.
The left and right panels show the CO-Solar and Def-Bound models, respectively.
The color scale indicates the terminal wind velocity, while the contours represent the degenerate core mass.
The red point with error bars denotes the observational estimate from \cite{2023ApJ...944..120L}.
\label{fg:HR_2models}}
\end{figure*}

In order to constrain the physical parameters of WD J005311, we compare our numerical results with the observational constraints.
Fig.~\ref{fg:HR_2models} shows the numerical line-driven wind solutions together with the observational constraints on the HR diagram.

\begin{figure*}[t]
\centering
\includegraphics[width=1.0\textwidth]{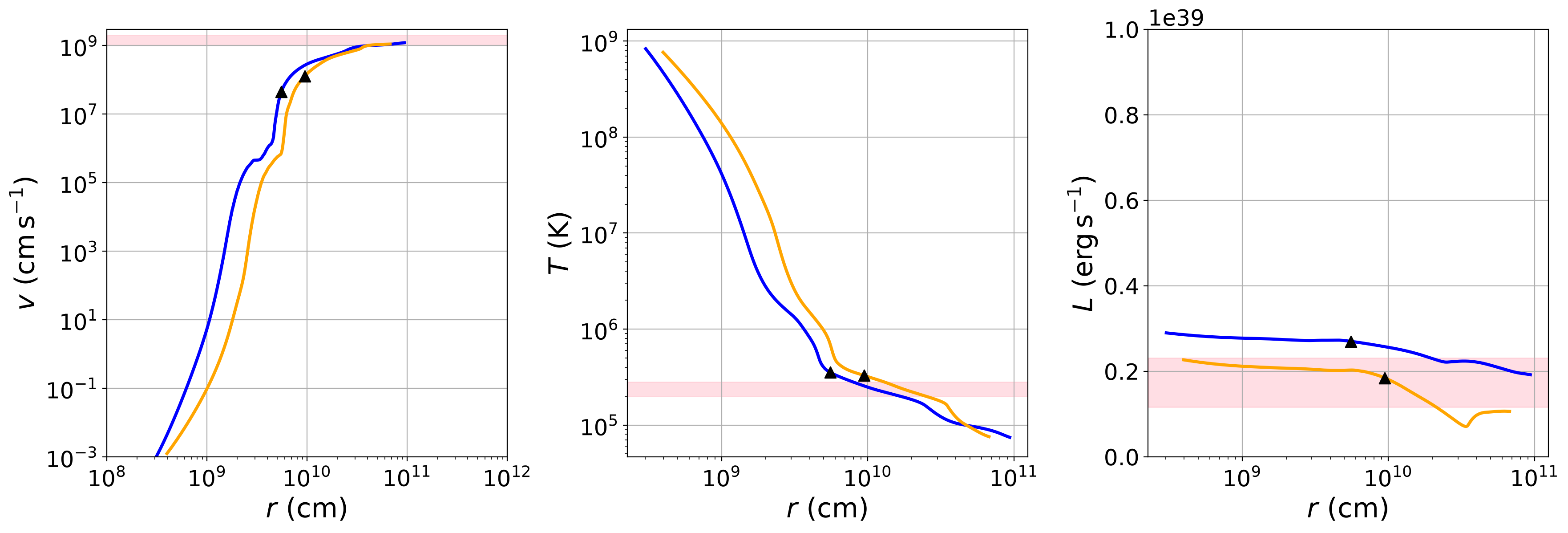}
\caption{
Two representative solutions that provide good overall agreement with the observational constraints for WD J005311.
The blue line represents the CO-Solar model with $M_{\rm WD}=1.25\, M_{\odot}$ and $\Delta M=1.8\times10^{-3}\, M_{\odot}$,
while the orange line represents the Def-Bound model with $M_{\rm WD}=1.15\, M_{\odot}$ and $\Delta M=5.7\times10^{-3}\, M_{\odot}$.
From right to left, the panels show the terminal wind velocity, effective temperature, and luminosity, respectively.
Triangle markers indicate the photospheric radius.
The shaded regions indicate the observational constraints for WD J005311.
The ranges of effective temperature and luminosity are adopted from
\cite{2023ApJ...944..120L},
while the velocity range corresponds to
$1.0\times10^{9}-2.0\times10^{9}\,{\rm cm\,s^{-1}}$.
\label{fg:bestfit_solutions}}
\end{figure*}

In both the CO-Solar and Def-Bound models, we find solutions that broadly satisfy the observational constraints on the terminal wind velocity, effective temperature, and luminosity.
Fig.~\ref{fg:bestfit_solutions} shows representative solutions for the CO-Solar and Def-Bound models that provide good overall agreement with the observational constraints.
The CO-Solar model has $M_{\rm WD}=1.25\, M_{\odot}$ and
$\Delta M=1.8\times10^{-3}\, M_{\odot}$,
while the Def-Bound model has $M_{\rm WD}=1.15\, M_{\odot}$ and
$\Delta M=5.7\times10^{-3}\, M_{\odot}$.

The difference in the required degenerate core mass between the two
chemical composition models can be attributed to their different
opacities.
The relatively low opacity of the CO-Solar model requires a higher
luminosity to drive the wind, resulting in a larger required degenerate
core mass.
In contrast, the higher opacity of the Def-Bound model allows the wind to
be driven with a smaller degenerate core mass.

The mass-loss rate of the CO-Solar model solution is
$\dot{M}=8.2\times10^{-7}\,M_{\odot}\,{\rm yr^{-1}}$,
while that of the Def-Bound model solution is
$\dot{M}=1.9\times10^{-6}\,M_{\odot}\,{\rm yr^{-1}}$.
Although the mass-loss rate of the Def-Bound model is closer to the observationally inferred value than that of the CO-Solar model \citep{2019Natur.569..684G},
both models give mass-loss rates that are broadly consistent with the observational constraint.

Despite the differences between the two chemical composition models,
both solutions share a common physical picture. 
The wind region contains only a very small amount of mass, 
indicating that the system is in the final stage of the optically thick wind phase. 
Therefore, WD J005311 can be interpreted as a merger remnant observed near the end of this phase.
As discussed in Sec.~\ref{ssec:timeevolution}, although the region of the $M_{\rm WD}-\Delta M$ plane that can reproduce the observational constraints is relatively small, 
the timescale over which the corresponding solution states persist accounts for a non negligible fraction of the total lifetime. 
Thus, observing WD J005311 in this late stage of the optically thick wind phase is not unlikely.

The total mass of the degenerate core and wind region in these solutions
is very large. It is difficult to form such a massive degenerate core in
only about $1000~{\rm yr}$, suggesting that the primary star of the
progenitor binary was likely a massive ONe WD.
However, the total mass is still below the Chandrasekhar mass limit.
Therefore, the merger remnant is not expected to undergo gravitational
collapse in the future.

\subsection{Time evolution of WD J005311}
As discussed in Sec.~\ref{ssec:timeevolution}, the time evolution of the
merger remnant can be reconstructed by connecting a sequence of wind
solutions. Using the representative solutions for the present-day
WD J005311 discussed in Sec.~\ref{ssec:constWDJ005311}, we can therefore
investigate the time evolution of the system since the merger.

\begin{figure}[t]
\centering
\includegraphics[width=0.4\textwidth]{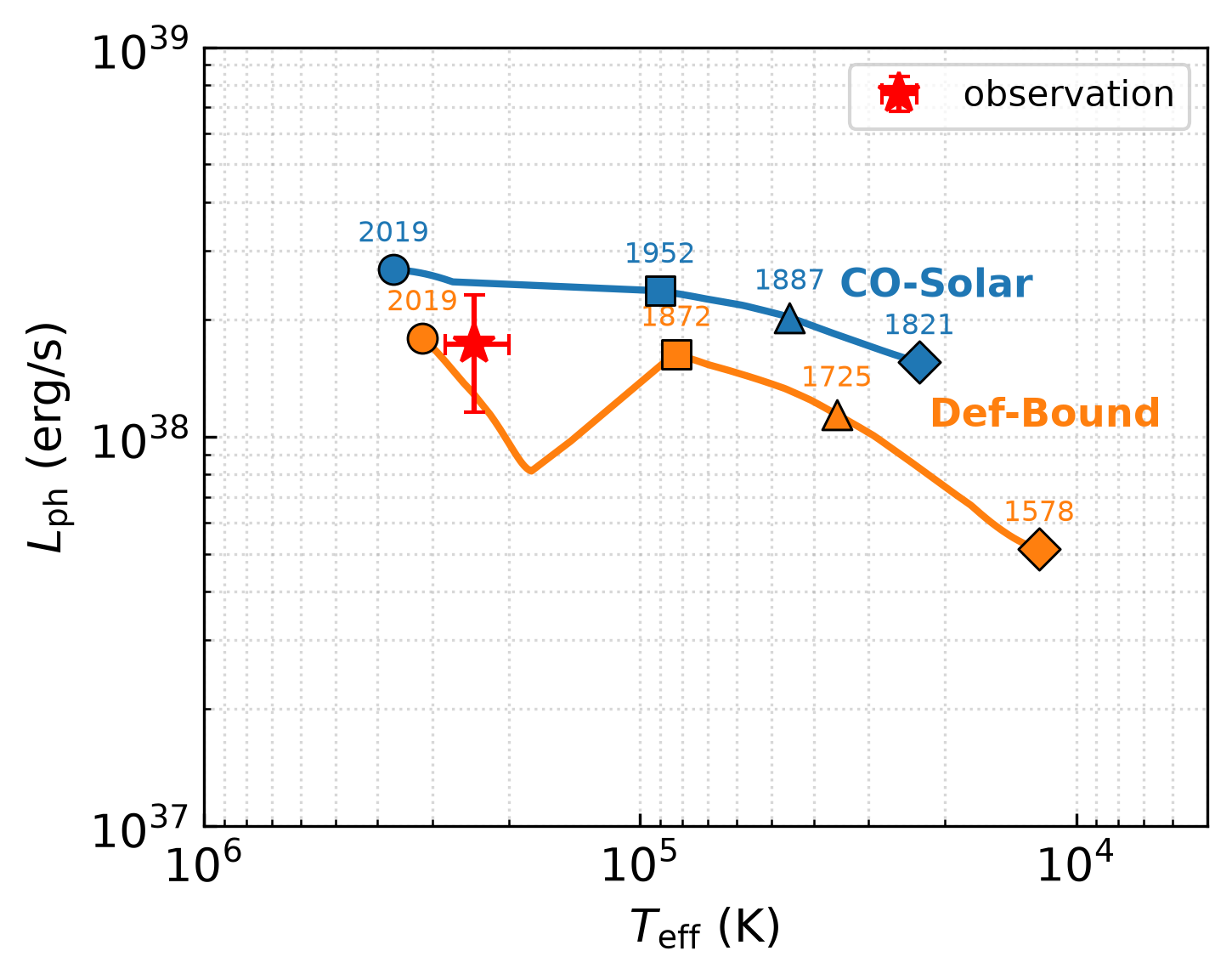}
\caption{
Time evolution of the system on the HR diagram, assuming that the two
representative solutions discussed in Sec.~\ref{ssec:constWDJ005311}
correspond to the current state of WD J005311.
The blue and orange tracks represent the CO-Solar and Def-Bound model
solutions, respectively.
Each marker indicates the calendar year of the evolutionary state on the HR diagram.
\label{fg:evoHR_WDJ005311}}
\end{figure}

Fig.~\ref{fg:evoHR_WDJ005311} shows the evolutionary tracks on the HR
diagram for the two representative solutions discussed in
Sec.~\ref{ssec:constWDJ005311}, assuming that each solution corresponds
to the current state of WD J005311.
The blue and orange tracks represent the CO-Solar and Def-Bound model
solutions, respectively. 

Assuming that the representative CO-Solar model
solution corresponds to the current state of WD J005311, the optically
thick wind phase has continued for approximately $200~{\rm yr}$ since
1823. During this period, the system evolves from right to left on the
HR diagram, with its luminosity remaining nearly constant.
On the other hand, assuming that the representative Def-Bound model solution
corresponds to the current state, the optically thick wind phase has
continued for approximately $450~{\rm yr}$ since 1580.

\cite{2023MNRAS.523.3885S} reported the long-term evolution of the
B-band photometry of WD J005311 based on Harvard College Observatory
photographic plates, covering approximately 150 years.
Their results show that the B-band brightness has gradually decreased
over a timescale of $\sim 100~{\rm yr}$.
The nearly horizontal evolutionary tracks shown in
Fig.~\ref{fg:evoHR_WDJ005311} may provide a possible explanation for
this long-term fading. Along these tracks, the luminosity remains
nearly constant while the effective temperature increases.
As the spectral energy distribution shifts toward higher frequencies,
the flux in the B band is expected to decrease.

However, a direct comparison between the simple theoretical B-band
magnitude evolution calculated from the effective temperature,
luminosity, and distance and the observed photometry is not appropriate.
The wavelength range of the B-band filter includes several emission
lines, such as ${\rm O~VI}$, and these contributions cannot be
accounted for by a simple blackbody approximation.
A detailed spectral calculation is therefore required for a quantitative
comparison with the observations.

Furthermore, from this evolutionary calculation, we find that
before 1821 for the CO-Solar model and before 1578 for the Def-Bound model,
the mass of the outer layer was too large to drive an optically thick wind.
We can predict that the system was in a hydrostatic giant phase,
in which the entire system expanded due to the massive envelope that
prevented the system from driving an optically thick wind, during the
period from the observed Type Iax supernova explosion in 1181, which
resulted from the merger, to 1821 or 1578, respectively.

Fiducially, the mass of the secondary CO WD is assumed to be about $0.6\,M_{\odot}$.
Since the mass of the outer nebula is estimated to be about $0.1\mbox{--}0.5\,M_{\odot}$ \citep{2020A&A...644L...8O, 2024ApJ...969..116K}, 
it is reasonable to expect that a mass of order $0.1\,M_{\odot}$ remains on the surface of the main ONe WD immediately after the merger.
Our results suggest that the mass of the outer layer decreases by an amount of order $0.1\,M_{\odot}$ during the hydrostatic giant phase over a timescale of a few hundred years,
either through mass loss or through stable shell burning that increases the core mass.

The relative importance of mass-loss and shell burning in reducing
the mass of the outer layer is not obvious.
Interestingly, the presence of a wind termination shock may provide
evidence for substantial mass-loss during the hydrostatic giant phase.
If this system indeed experienced a hydrostatic giant phase, the wind termination
shock identified by X-ray emission may have been produced by the
collision between the circumstellar material (CSM) formed during the hydrostatic giant phase and the
subsequent wind during the optically thick wind phase. Given the
timescale of the optically thick wind phase and the kinetic energy
injected by the wind during this phase, we estimate that a CSM
mass of order $0.1\,M_{\odot}$ is sufficient to account for the
observed location of the wind termination shock. In this scenario,
the shock can be explained without requiring the onset of the
optically thick wind phase to be fine-tuned to about 10 yr ago, as
suggested by \cite{2024ApJ...969..116K}. Instead, the required CSM can be
naturally formed through intense mass-loss during the preceding hydrostatic giant
phase. The wind termination shock may therefore provide indirect
evidence for substantial mass-loss during the hydrostatic giant phase.

\begin{figure*}[t]
\centering
\includegraphics[width=0.9\textwidth]{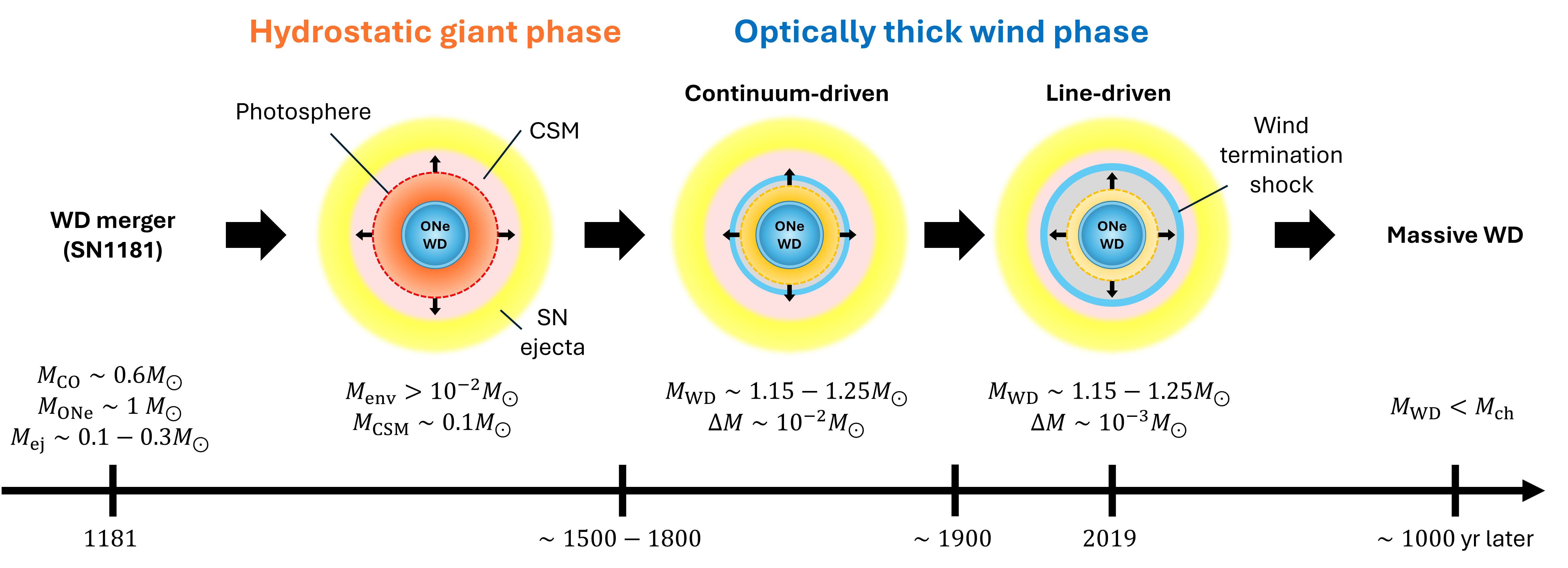}
\caption{
Schematic illustration of the evolution of WD J005311
from SN 1181 to the future.
\label{fg:WDJ005311_evo_cartoon}}
\end{figure*}

Finally, we can predict the future evolution of this object.
The optically thick wind phase will continue for
$\sim \Delta M/\dot{M}\sim 1000~{\rm yr}$.
Since the total mass of the remnant is below the Chandrasekhar
mass limit, the object is expected to evolve into a massive,
cool WD after the optically thick wind phase.
This evolutionary scenario is illustrated in Fig.~\ref{fg:WDJ005311_evo_cartoon}.

\section{Conclusion}
In this paper, we investigated optically thick wind solutions of WD merger remnants
in various evolutionary states,
with one of our main goals being to trace the physical state and evolution of WD J005311.
We obtained the following general insights into WD merger remnants.
\begin{itemize}
    \item
    In systems with a degenerate core mass larger than $1.0\,M_{\odot}$,
    optically thick winds, including continuum-driven wind and line-driven wind,
    are realized for a certain range of wind region masses.
    \item
    For a relatively large wind region mass,
    continuum absorption dominates the wind acceleration,
    and a relatively slow continuum-driven wind solution is obtained.
    As the wind region mass decreases, absorption by spectral lines becomes increasingly important,
    resulting in a relatively fast line-driven wind solution.
    \item
    For larger degenerate core masses,
    the range of wind region masses for which optically thick winds can be driven
    shifts toward smaller values.
    At the same time, the maximum terminal wind velocity
    attained within this range is higher.
    \item 
    The timescale of the optically thick wind phase
    is shorter for larger degenerate core masses,
    ranging from $\sim 10^3~{\rm yr}$ for $1.05\,M_{\odot}$
    to $\sim 10~{\rm yr}$ for $1.35\,M_{\odot}$.
    \item 
    The line-driven wind phase is realized during the later stage of the
    optically thick wind phase, characterized by a high terminal wind velocity,
    and lasts for a relatively long time because of the low mass-loss rate.
\end{itemize}

We identified wind solutions from the sets of solutions obtained
for the two chemical composition models that can reproduce the observed
terminal wind velocity, effective temperature, and luminosity of WD J005311.
The following evolutionary implications can be drawn from these results.
\begin{itemize}
    \item
    The line-driven wind solution with a degenerate core mass of $1.25\,M_{\odot}$
    and a wind region mass of $1.8\times10^{-3}\,M_{\odot}$
    for the CO-Solar model,
    and the line-driven wind solution with a degenerate core mass of $1.15\,M_{\odot}$
    and a wind region mass of $5.7\times10^{-3}\,M_{\odot}$
    for the Def-Bound model,
    broadly reproduce the observational properties of WD J005311.
    In both cases, these solutions correspond to the late stage of the optically thick wind phase.
    \item 
    Based on the evolutionary analysis using steady-state solutions,
    WD J005311 is inferred to have experienced a hydrostatic giant phase lasting
    about 400--650 yr after the explosion,
    followed by a continuum-driven wind phase lasting
    about 100--350 yr and a line-driven wind phase lasting
    about 100 yr.
    \item 
    Taken together, our evolutionary scenario and the observed wind termination shock suggest 
    that the remnant may have undergone substantial mass loss of $\sim 0.1\,M_{\odot}$ 
    during the relatively short 400--650 yr hydrostatic giant phase.
    \item 
    This optically thick wind phase is expected to continue
    for $\sim 1000~{\rm yr}$ into the future,
    after which the remnant will become a massive, cool WD.
\end{itemize}

Taken together, these results have general implications for the fate of WD merger remnants.
If the progenitor of WD J005311 was a fiducial WD binary,
its total mass was likely super-Chandrasekhar.
Nevertheless, the degenerate core mass inferred in this study
is sub-Chandrasekhar.
This suggests that even a merger of a CO and an ONe WD
may not necessarily leave a remnant capable of undergoing core collapse,
possibly because of substantial mass-loss during the hydrostatic giant phase.

If energy injection into the CSM by the wind continues,
the position of the wind termination shock is expected to change
on a timescale of $\sim 10$--$100~{\rm yr}$,
based on the energy injection rate.
Therefore, by combining future observations of the wind nebula
with the time evolution of the energy injected by the wind
predicted by our calculations,
it may be possible to constrain the evolution of the system
during the hydrostatic giant phase.
A detailed calculation of the global evolution of a
double-degenerate WD merger remnant, including the hydrostatic giant phase,
will be an important subject for future work.

\begin{acknowledgments}
We thank Kotaro Fujisawa, Kengo Tomida, Kazuyuki Omukai,
Yuki Kudoh, Toshikazu Shigeyama, Daichi Tsuna, and Wataru Ishizaki
for valuable discussions, comments, and suggestions.


The computations in this work are partly performed on Resceubbc 
at the Research Center for the Early Universe, The University of Tokyo.

This research was partially supported by JSPS KAKENHI Grant
Nos.~23H04899, 24K00668, and 25K00021 (KK).
\end{acknowledgments}

\appendix

\section{Calculation of lineforce multiplier} \label{Asec:linemult}
\subsection{Calculation of the Level Population} 
\label{Assec:levpop}
We assume that the fluid is in local thermodynamic equilibrium (LTE)
with temperature $T$ and density $\rho$.
In order to calculate $M(t)$,
it is necessary to evaluate the number density $n^{\rm el}_{i,j}$
for a given chemical composition, temperature, and density.
This quantity can be written as
\begin{equation}
    n^{\rm el}_{i,j}
    =
    n^{\rm el}_i \frac{n^{\rm el}_{i,j}}{n^{\rm el}_i} , \label{eq:densityratej}
\end{equation}where $n^{\rm el}_i$ represents the number density of species ``${\rm el}$'' in ionization stage $i$.

The first factor on the right-hand side of Eq.~\ref{eq:densityratej}
is determined by the Saha equation for all elements and ionization states.
The Saha equation is
\begin{equation}
    \frac{n^{\rm el}_{i+1}}{n^{\rm el}_{i}} n_{\rm e}
    =
    \frac{2}{\lambda_{\rm e}^3} \frac{U^{\rm el}_{i+1}(T)}{U^{\rm el}_{i}(T)} \exp{\left(- \frac{\chi^{\rm el}_{i}}{k_{\rm B} T}\right)},
\end{equation}where $U^{\rm el}_{i}$ and $\chi^{\rm el}_{i}$ are the partition function and ionization energy, respectively, for species ``${\rm el}$'' in ionization stage $i$, and $\lambda_{\rm e}$ is the de Broglie wavelength of an electron,
\begin{equation}
    \lambda_{\rm e}
    =
    \frac{h}{\sqrt{2\pi m_{\rm e} k_{\rm B} T}}.
\end{equation}

The partition function is defined as
\begin{equation}
    U^{\rm el}_i(T)
    =
    \sum_{j}
    g^{\rm el}_{i,j}
    \exp\left[
    -\frac{E^{\rm el}_{i,j}-E^{\rm el}_{i,0}}{k_{\rm B}T}
    \right],
\end{equation}where $g^{\rm el}_{i,j}$ is the statistical weight of level $j$ of ionization stage $i$ of species ``${\rm el}$'',
and $E^{\rm el}_{i,j}-E^{\rm el}_{i,0}$ is the excitation energy from the ground state.
Following \citet{2010ApJ...711..239C},
we adopt the approximate expression
\begin{equation}
    U^{\rm el}_i(T)
    \simeq
    g^{\rm el}_{i,0}
    +
    G^{\rm el}_{i}
    \exp\left[
    -\frac{\epsilon^{\rm el}_{i}}{k_{\rm B}T}
    \right]
    +
    \frac{m^{\rm el}_{i}}{3}
    \left(
    {n^{\rm el}_{*,i}}^3-343
    \right)
    \exp\left[
    -\frac{E^{\rm el}_{n_*i}}{k_{\rm B}T}
    \right],
\end{equation}where $g^{\rm el}_{i,0}$ is the ground-state statistical weight,
and $E^{\rm el}_{n_*i}$ is given by
\begin{equation}
    E^{\rm el}_{n_* i}
    =
    \chi^{\rm el}_{i}
    -
    \frac{{Z_{\rm eff}}^2 E_{\rm Ry}}{{n^{\rm el}_{*,i}}^2},
\end{equation}for ionization stage $i$.
Here, $\chi^{\rm el}_i$ is the ionization potential,
$E_{\rm Ry}$ is the Rydberg energy,
and $Z_{\rm eff}$ is the effective ionic charge, given by $(i+1)$.
The quantity $n^{\rm el}_{*,i}$ is defined as
\begin{equation}
    n^{\rm el}_{*,i}
    =
    \frac{q^{\rm el}_{i}}{2}
    \left(
    1+\sqrt{1+\frac{4}{q^{\rm el}_{i}}}
    \right),
\end{equation}with
\begin{equation}
    q^{\rm el}_{i}
    =
    \sqrt{
    \frac{Z_{\rm eff}}{2\pi a_0}
    }
    {n_{\rm tot}}^{-\frac{1}{6}},
\end{equation}where $a_0$ is the Bohr radius.
The total number density of the gas is given by
\begin{equation}
    n_{\rm tot}
    =
    n_{\rm e}
    +
    \sum_{\rm el}
    \sum_i n^{\rm el}_i.
\end{equation}The fitting parameters $G^{\rm el}_i$, $\epsilon^{\rm el}_i$, and $m^{\rm el}_{i}$
are taken from the tables of \citet{2010ApJ...711..239C}.
However, these tabulated values are available only for elements with $Z<20$.
For heavier elements ($Z>20$),
we adopt the empirical fitting relations introduced by
\citet{2021AAS...23711602L},
\begin{equation}
    \epsilon^{\rm el}_{i}
    =
    \chi^{\rm el}_i
    \left(
    0.946-0.007Z_{\rm eff}
    \right),
\end{equation}
\begin{equation}
    m^{\rm el}_{i}
    =
    4.0
    g_0^{0.79},
\end{equation}and
\begin{equation}
    G^{\rm el}_i
    =
    113m^{0.66}.
\end{equation}The ionization potentials $\chi^{\rm el}_i$
and ground-state statistical weights $g^{\rm el}_{i,0}$
are taken from the National Institute of Standards and Technology database (NIST; \citeauthor{2018APS..DMPM01004K} \citeyear{2018APS..DMPM01004K}).

To close the system of equations, we require additional constraints in the form of number-density conservation equations.
These constraints are provided by the number-density conservation equations,
\begin{equation}
\frac{\rho X_{\rm el}}{m_{\rm el}}
=
\sum_i n^{\rm el}_{i},
\end{equation}where $X_{\rm el}$ is the mass fraction of species ``${\rm el}$''
and $m_{\rm el}$ is the mass of a particle of species ``${\rm el}$'',
together with the electron number-density conservation equation,
\begin{equation}
n_{\rm e}
=
\sum_{\rm el} \sum_i n^{\rm el}_i i.
\end{equation}Together with the Saha equations, these conservation equations
completely determine all ionization-state number densities and 
the electron number density.

\begin{figure*}[t]
\centering
\includegraphics[width=0.4\textwidth]{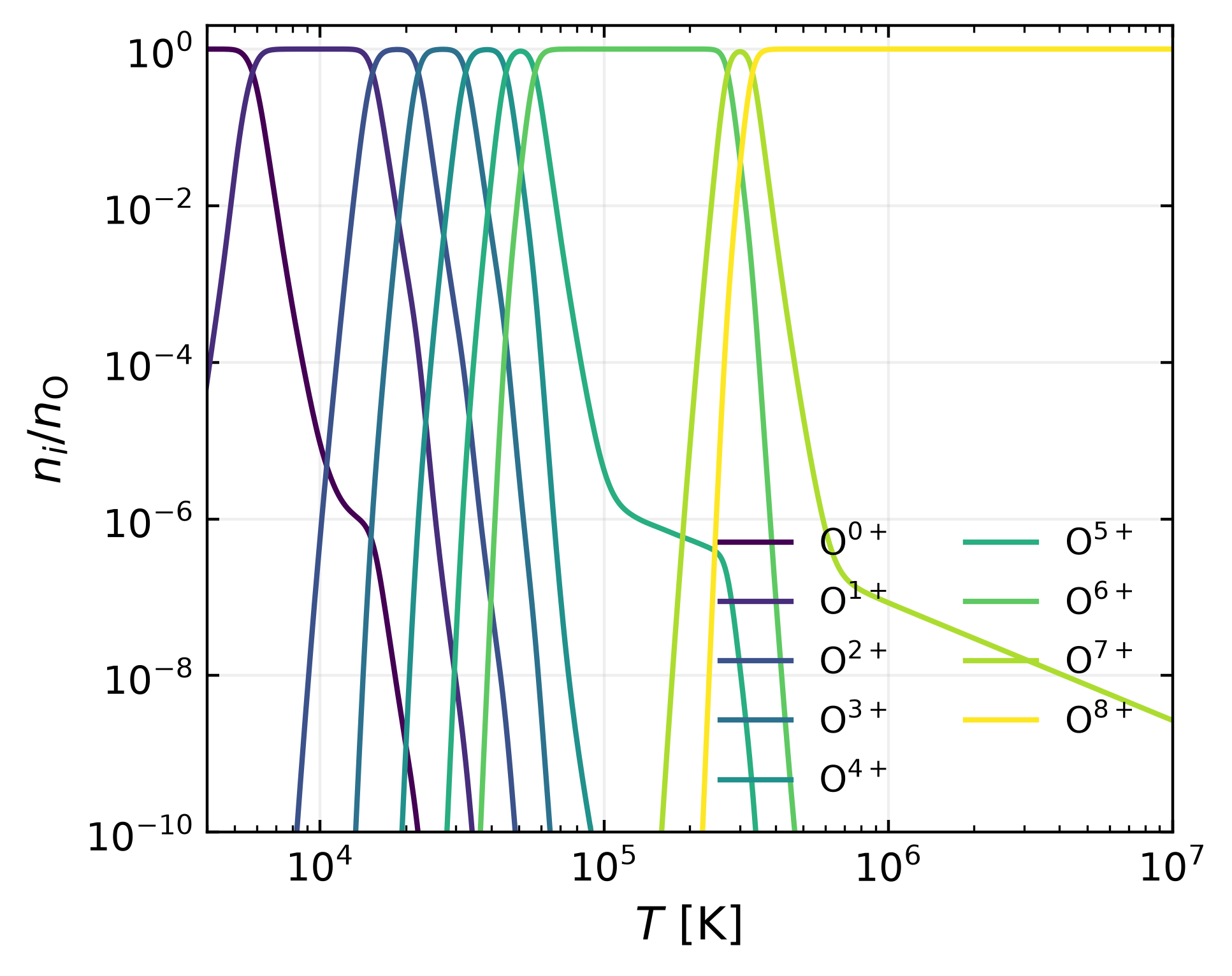}
\caption{
Ionization-state transitions of oxygen as functions of temperature,
calculated from ionization equilibrium under the assumption of LTE.
Each line represents an individual ionization state.
The chemical composition is taken to be the CO-Solar model in Table~\ref{tb:compmodels},
and the density is fixed at
$\rho = 1.0 \times 10^{-13}~{\rm g~cm^{-3}}$.
\label{fg:ionO}}
\end{figure*}

In order to solve these equations self-consistently, we employ an iterative method.
As an initial solution, the number densities of all ionization states of all elements are set to equal values.
The solution is then iterated until the residuals become sufficiently small.
Fig.~\ref{fg:ionO}
shows the calculated ionization states of oxygen
as functions of temperature
for a fixed chemical composition and density.

In preparation for calculating the shape of the function $M(t)$
for a given chemical composition over a range of temperatures and densities,
we construct a grid in temperature and density while keeping the composition fixed.
For each grid point, the ionization fractions of all elements are calculated and stored in a lookup table.

The second factor on the right-hand side of Eq.~\ref{eq:densityratej} represents
the population ratio of excited level $j$
to the total population of ionization state $i$.
Under LTE, this ratio is given by the Boltzmann distribution,
\begin{equation}
    \frac{n^{\rm el}_{i,j}}{n^{\rm el}_i}
    =
    \frac{
    g^{\rm el}_{i,j}
    \exp\left[
    -\frac{E^{\rm el}_{i,j}-E^{\rm el}_{i,0}}{k_{\rm B}T}
    \right]
    }{
    U^{\rm el}_i(T)
    }. \label{eq:jstaterate}
\end{equation}Using these results together with the solutions of the Saha equations,
the level populations can be determined.
Note that the excitation energies and statistical weights are taken
from the atomic line data used in the line-force calculations.
Details of the adopted atomic data are described in the next section.

\subsection{Numerical Calculation of the Line-Force Multiplier} \label{Assec:Mtcalc}
For a given chemical composition, we construct the form of $M(t)$
over a range of temperatures and densities.
To do so, we must sum the contributions of all bound--bound transitions
from all elements according to Eq.~\ref{eq:forcemult}.

To ensure the most complete possible set of bound--bound transitions,
we compiled a line list by combining several publicly available atomic
databases, including NIST \citep{2018APS..DMPM01004K}, version 11.0.2
of the CHIANTI atomic database \citep{2024ApJ...974...71D}, the atomic
database used by the radiative transfer code CMFGEN
\citep{1998ApJ...496..407H}, and AtomDB version 3.1.3 accessed through
PyAtomDB version 1.0.2 \citep{2012ApJ...756..128F}. For each ion, we
adopt the database providing the most complete atomic data. The
oscillator strengths, transition energies, statistical weights, and
lower-level energies are then extracted from the resulting line list.
The final line list contains 102,982,215.

\begin{figure*}[t]
\centering
\includegraphics[width=0.4\textwidth]{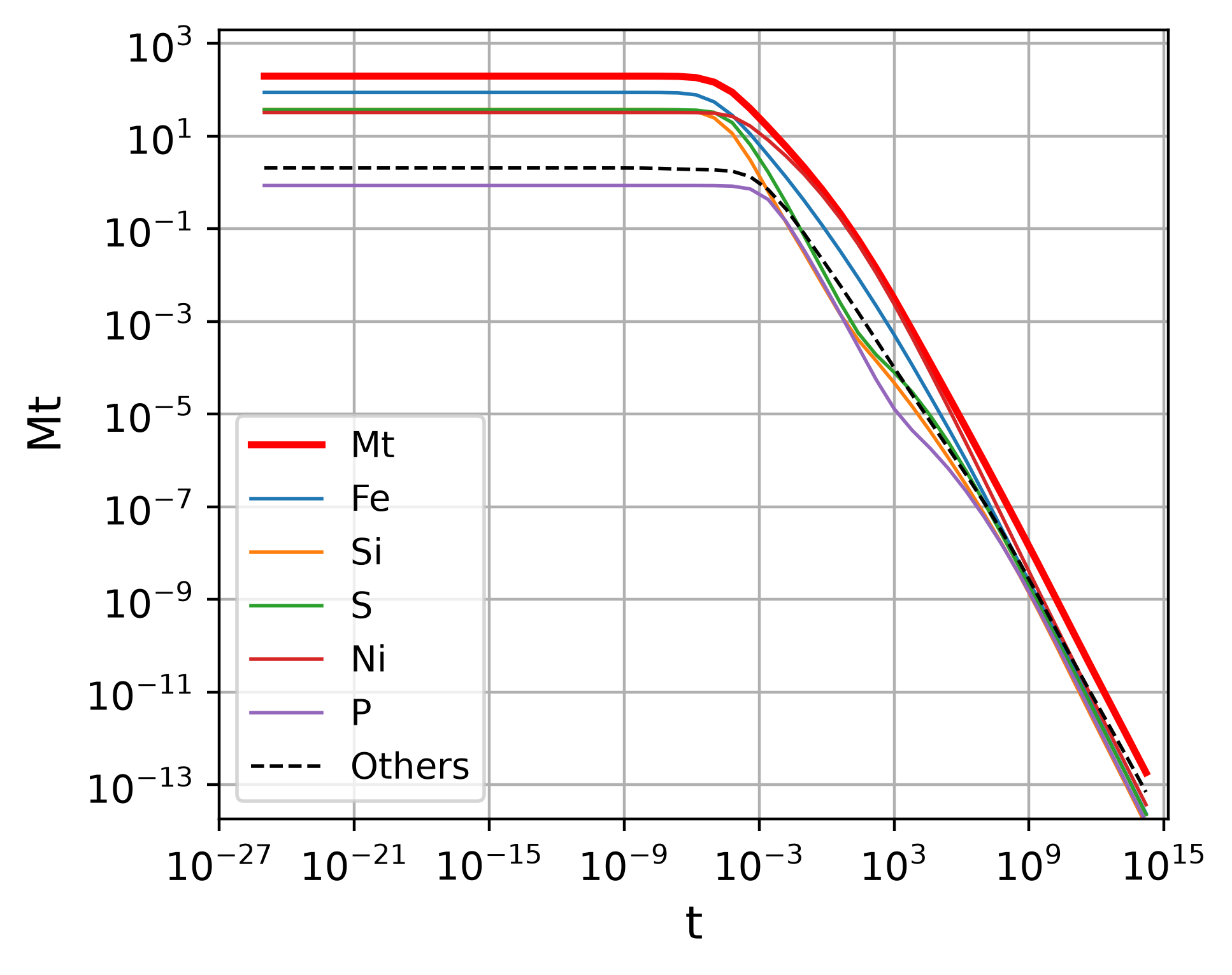}
\caption{
Line-force multiplier calculated from the line data.
The black curve shows the contribution from all lines,
while the colored curves show the contributions from individual elements.
The chemical composition is assumed to be the CO-Solar model in Table~\ref{tb:compmodels}.
The density and temperature are fixed at $\rho = 1.0 \times 10^{-11}~{\rm g~cm^{-3}}$ 
and $T = 2.0 \times 10^{5}~{\rm K}$, respectively.
\label{fg:Mt}}
\end{figure*}

For a given temperature, density, and value of $t$, the force multiplier
$M(t)$ is calculated using the ionization fractions obtained from the
lookup table together with the atomic data extracted from the line list,
including oscillator strengths, transition energies, and lower-level
energies. This calculation is repeated for 15 logarithmically spaced
values of $t$ between $1\times10^{-18}$ and $1\times10^{32}$, thereby
determining the shape of $M(t)$ for each temperature--density pair.
Fig.~\ref{fg:Mt} shows an example of $M(t)$ for a specific density, temperature, and chemical composition.

\subsection{Parameterization of $M(t)$}\label{Assec:Mtfit}
\begin{figure*}[t]
\centering
\includegraphics[width=1.0\textwidth]{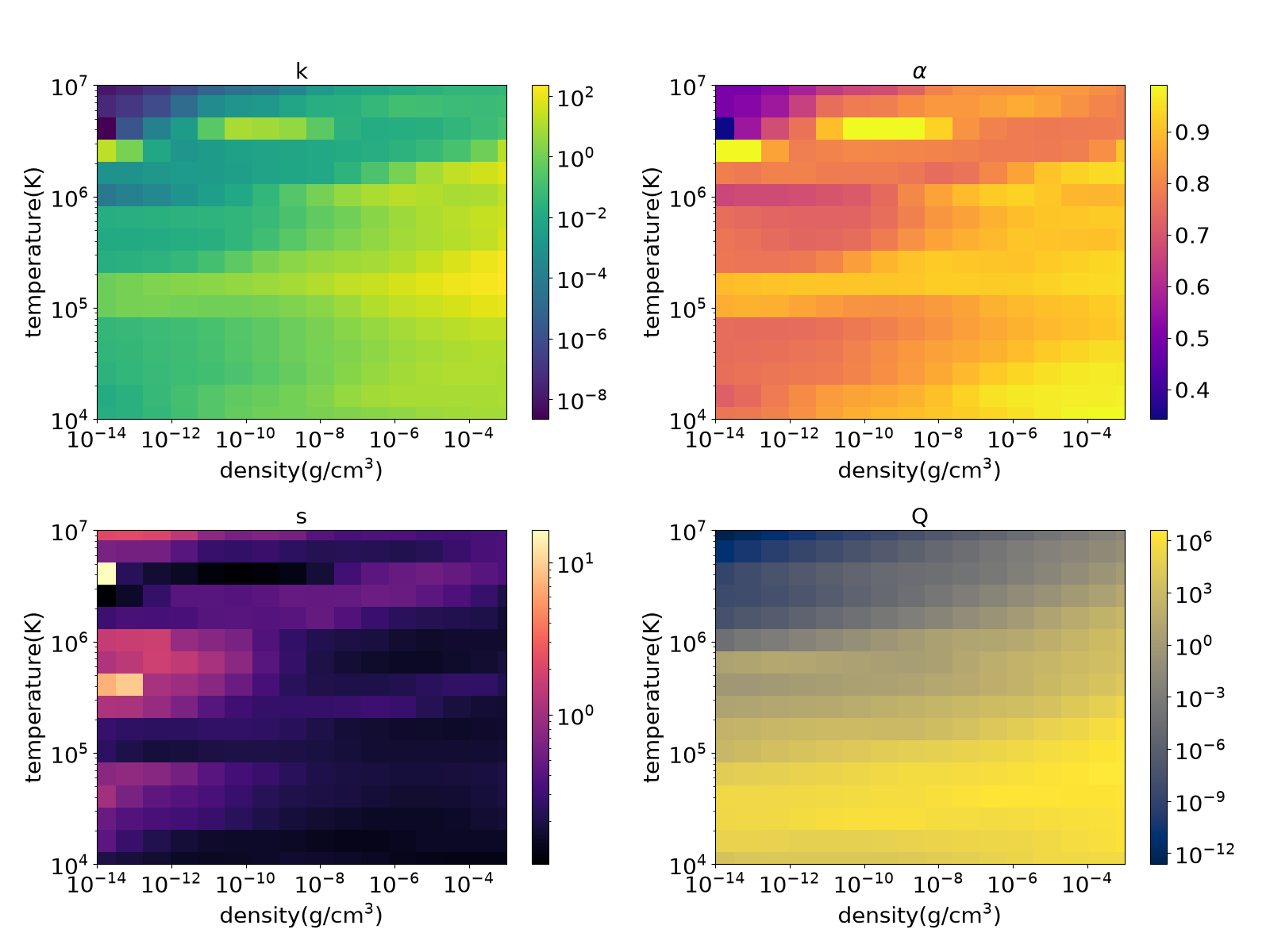}
\caption{
Distribution of the fitting parameters
$k$, $\alpha$, $s$, and $\bar{Q}$,
obtained by fitting the approximate expression of $M(t)$
at each temperature and density
for a fixed chemical composition.
The chemical composition is taken to be the CO-Solar model in Table~\ref{tb:compmodels}.
\label{fg:4params}}
\end{figure*}

Following \cite{2021AAS...23711602L},
the line-force multiplier is approximated by
\begin{equation}
    M(t)
    =
    \frac{
    \bar{Q} k
    }{
    \left(
    k^s + \bar{Q}^s t^{\alpha s}
    \right)^{\frac{1}{s}}
    }.
\end{equation}Here, $k$, $\alpha$, $s$, and $\bar{Q}$ are fitting parameters.
Using this functional form, we perform nonlinear least-squares fits to the numerically calculated $M(t)$.
This procedure is repeated for each temperature--density pair,
yielding fitting parameters that are tabulated as functions of
temperature and density for each fixed chemical composition.
Fig.~(\ref{fg:4params}) shows fitting results of 4 parameters. 
Using these parameters,
the contribution of the line force can be calculated at any point in the wind.

\section{Iterative Method for Solving wind solution} \label{Asec:itemet}
\subsection{Definition of Numerical Variables}
The system consists of four equations: the Euler equation
(Eq.~\ref{eq:Euler}), the energy transport equation
(Eq.~\ref{eq:effdiffusion}), the continuity equation
(Eq.~\ref{eq:cont}), and the energy conservation equation
(Eq.~\ref{eq:energycons}).
In addition, the solution must satisfy conditions at three characteristic locations:
the inner boundary, the outer boundary, and the singular point.
These equations are solved self-consistently for the dependent variables
$v$, $T$, $\rho$, and $L$ as functions of the independent variable $r$.

However, by introducing the integration constants
$\dot{M}$ and $\Lambda_{\rm tot}$,
the continuity equation and the energy conservation equation
can be written in integral form.
The density $\rho$ and luminosity $L$ can then be expressed
in terms of $\dot{M}$, $\Lambda_{\rm tot}$, $r$, $v$, and $T$.
As a result, the dependent variables are reduced to
$v$ and $T$.

In addition, the locations of the inner boundary, the singular point, and the photosphere are not known a priori and must be determined as part of the solution.
These radii therefore constitute additional unknown constants.
Consequently, for a given chemical composition,
$M_{\rm WD}$, and $\Lambda_{\rm tot}$,
the problem reduces to solving for the dependent variables
$v(r)$ and $T(r)$ together with the four unknown constants
$\dot{M}$, $r_{\rm in}$, $r_{\rm s}$, and $r_{\rm ph}$.

On the other hand, when solving this boundary-value problem numerically using the relaxation method, it is necessary to carefully reformulate the treatment of both dependent and independent variables.
The iterative method solves a boundary-value problem by discretizing the differential equations on a numerical grid. 
Starting from an initial trial solution, the residuals are evaluated and the Jacobian matrix is constructed.
The Jacobian matrix is used to determine the first-order corrections to the solution.
The trial solution is updated with these corrections, and the procedure is repeated until the corrections become sufficiently small.

In order to evaluate the Jacobian matrix, the mesh indices corresponding to the inner boundary, the outer boundary, and the singular point must be specified in advance.
However, as discussed above, the physical locations of these points are not known a priori and can only be determined as part of the solution.
Therefore, the radial coordinate cannot be assigned to each mesh point before the solution is obtained.
For this reason, we treat also the radial coordinate, $r$, as a dependent variable and solve for it as a function of the mesh index $q$.

Also, a common approach in the relaxation method is to treat unknown constants as dependent variables of the mesh coordinate $q$ and to introduce trivial differential equations for them.
In this formulation, the unknown constants are determined simultaneously with the other variables as part of the relaxation procedure.
We adopt this approach and treat $r_{\rm in}$, $r_{\rm s}$, $r_{\rm ph}$, and $\dot{M}$ as dependent variables of $q$.

In this paper, the chemical composition, $M_{\rm WD}$, and $\Lambda_{\rm tot}$ are treated as given parameters.
Accordingly, we treat $\log_{10}{r(q)}$, $\log_{10}{v(q)}$, $\log_{10}{T(q)}$, $\log_{10}{r_{\rm s}(q)}$, $\log_{10}{r_{\rm ph}(q)}$, and $\log_{10}{\dot{M}(q)}$ as dependent variables of the mesh coordinate $q$.
Note that $r_{\rm in}$ is excluded because it is uniquely determined by the specified value of $M_{\rm WD}$ and therefore need not be treated as a dependent variable.
Since these quantities vary over several orders of magnitude across the domain, we solve the equations in logarithmic form for numerical stability.

\subsubsection{The Differential Equations}
For $\log_{10} r$, we define its differential equation such that specific mesh points correspond to the inner boundary, the singular point, and the photosphere.
Therefore, we prescribe the following form:
\begin{equation}
   \frac{d \log_{10} r}{d q} 
   = 
   \begin{cases}
       \dfrac{\log_{10} r_{\rm s}- \log_{10} r_{\rm in}}{N_{\rm s} - 1} & (0 \leq q < N_{\rm s}) \\
       \dfrac{\log_{10} r_{\rm ph}- \log_{10} r_{\rm s}}{N - N_{\rm s}} & (N_{\rm s} \leq q < N)
   \end{cases},
\end{equation}where $N$ is the total number of mesh points, and $N_{\rm s}-1$ denotes the mesh index of the singular point.

For $\log_{10} v$, the governing differential equation is the Euler equation (Eq.~\ref{eq:Euler}).
Taking into account the dependence of $\kappa_{\rm eff}$ on $dv/dr$, the Euler equation can be written as follows,
\begin{equation}
    F \left(x \equiv \frac{d \log_{10} v}{d \log_{10} r}\right)
    \equiv 
    x +
    \Gamma_{1}+\Gamma_{2}\left(\Gamma_{3} +
    x^{-\alpha s}\right)^{-\frac{1 }{s }} 
    =
    0,
\end{equation}with
\begin{equation}
    \Gamma_{1} 
    \equiv
    \frac{-2 + \frac{GM_{\rm WD}}{c_s^2 r }-\frac{3\rho \kappa_{\rm R} L}{16\pi r c a T^4}\left(1+\frac{4aT^4 }{3 \rho c_s^2 }\right)}{\frac{v^2 }{c_s^2 }-1},
\end{equation}
\begin{equation}
    \Gamma_{2} 
    \equiv
    \frac{-\frac{3\rho k \kappa_{\rm es,f} L}{16\pi r c a T^4}\left(1+\frac{4aT^4 }{3 \rho c_s^2 }\right)\left(\frac{v }{\kappa_{\rm es,f}\rho r c_s}\right)^{\alpha}}{\frac{v^2 }{c_s^2 }-1},
\end{equation}
\begin{equation}
    \Gamma_{3} 
    \equiv
    \left(\frac{k}{\bar{Q}} \left(\frac{v }{\kappa_{\rm es,f}\rho r c_s }\right)^{\alpha}\right)^s.
\end{equation}The quantities $\Gamma_1$, $\Gamma_2$, and $\Gamma_3$ are functions of the physical variables $r$, $v$, $T$, $\rho$, and $L$.
The parameters $\alpha$ and $s$ are determined by $T$ and $\rho$.
Depending on these quantities, the number of real solutions and the presence of local extrema vary.

Table~\ref{tb:numofsol} summarizes the number of solutions of the Euler equation for different combinations of $\Gamma_{1}$ and $\Gamma_{2}$.
Here, $x_{\rm lmin}$ means $d\log_{10}v/d\log_{10}r$ at the local extrema,
$C_{\rm i}$ and $C_{\rm o}$ branches represent the two solution branches of the Euler equation.
When tracking solutions across different regions,
solutions belonging to the same branch connect continuously,
whereas solutions belonging to different branches cannot be connected continuously.
\begin{deluxetable}{cccc}
\tablecaption{Number of solutions of the Euler equation
\label{tb:numofsol}}
\tablehead{
\colhead{Region} & \colhead{Conditions} & \colhead{$C_{\rm i}$ branch} & \colhead{$C_{\rm o}$ branch}
}
\startdata
(I) & $\Gamma_1 < 0,\ \Gamma_2 > 0$ & 1 & 0 \\
(I\hspace{-1.2pt}I) & $\Gamma_1 > 0,\ \Gamma_2 < 0,\ F(x_{\rm lmin}) < 0$ & 1 & 1 \\
(I\hspace{-1.2pt}I\hspace{-1.2pt}I) & $\Gamma_1 > 0,\ \Gamma_2 < 0,\ F(x_{\rm lmin}) > 0$ & 0 & 0 \\
(I\hspace{-1.2pt}V) & $\Gamma_1 > 0,\ \Gamma_2 > 0$ & 0 & 0 \\
(V) & $\Gamma_1 < 0,\ \Gamma_2 < 0$ & 0 & 1
\enddata
\end{deluxetable}
\onecolumngrid

At the inner boundary, the physical state corresponds to region (I), and thus the solution of the Euler equation follows the $C_{\rm i}$ branch.
On the other hand, at the outer boundary, the physical state corresponds to region (V), and thus the solution of the Euler equation follows the $C_{\rm o}$ branch.
Therefore, in order to connect the velocity gradient from the inner boundary to the outer boundary, the solution branch must transition from one branch to the other at a specific point.
This point corresponds to the singular point, where both the singular condition and the regularity condition must be satisfied.

When numerically constructing a steady-state solution,
we adopt the $C_{\rm i}$ branch as the solution of the differential equation for $\log_{10} v$ at mesh points interior to the singular point.
Conversely, at mesh points exterior to the singular point, we adopt the $C_{\rm o}$ branch.

The singular condition requires that the Euler equation have a double root.
Therefore, the solution must lie on the boundary between region (I\hspace{-1.2pt}I)
and region (I\hspace{-1.2pt}I\hspace{-1.2pt}I), 
where the two real solution branches merge and cease to exist.
During the iterative procedure, the trial solution approaches this boundary progressively.
As a result, it is not uncommon for an intermediate trial solution to enter region (I\hspace{-1.2pt}I\hspace{-1.2pt}I), 
where the Euler equation becomes undefined,
making it impossible to evaluate the residuals required for the subsequent iteration step.
On the other hand, if entry into region (I\hspace{-1.2pt}I\hspace{-1.2pt}I) were strictly prohibited during the iterative procedure, 
the convergence properties of the method would be significantly degraded.

To alleviate this difficulty,
we introduce an artificial differential equation in region (I\hspace{-1.2pt}I\hspace{-1.2pt}I)
during the iterative procedure, allowing intermediate trial solutions to traverse this region.
To ensure good convergence,
the solutions of the artificial differential equation must connect smoothly to those of the original differential equation in region (I\hspace{-1.2pt}I).
We therefore adopt the following equation as the artificial differential equation in region (I\hspace{-1.2pt}I\hspace{-1.2pt}I), which guarantees this continuity,
\begin{equation}
    F_{\rm (I\hspace{-1.2pt}I\hspace{-1.2pt}I)}
    \equiv
    -\left(x-x_{\rm lmin}\right)^2 - \frac{x}{x_{\rm lmin}}
    -\frac{1}{x} + F\left(x_{\rm lmin}\right) + \frac{2}{x_{\rm lmin}}
    = 0.
\end{equation}We define the solution branch that connects smoothly to the $C_{\rm i}$ branch in region (\mbox{I\hspace{-1.2pt}I}) as the $C_{\rm i}$ branch in region (\mbox{I\hspace{-1.2pt}I\hspace{-1.2pt}I}).
Likewise, we define the solution branch that connects smoothly to the $C_{\rm o}$ branch in region (\mbox{I\hspace{-1.2pt}I}) as the $C_{\rm o}$ branch in region (\mbox{I\hspace{-1.2pt}I\hspace{-1.2pt}I}).

The solution of the differential equation for $\log_{10} v$ as a function of $q$ is then obtained by combining the solution obtained above with the differential equation governing $\log_{10} r$ as a function of $q$.
After convergence is achieved, we check whether the converged solution enters region (\mbox{I\hspace{-1.2pt}I\hspace{-1.2pt}I}).
If it does not, the solution is accepted as a valid solution of the original differential equation.

Also, if the Rosseland mean opacity becomes sufficiently large in the region
interior to the singular point, the solution may enter region
(\mbox{I\hspace{-1.2pt}V}) before reaching the singular point.
Since no $C_{\rm i}$-branch solution exists in this region,
we instead adopt the artificial condition
$d \ln v / d \ln r = 0$.
This treatment reflects an intrinsic limitation of the present formulation
including the line force, which cannot describe a continuously connected
solution with a negative velocity gradient.
In practice, the region where this artificial treatment is applied occupies
only a very small fraction of the entire solution and does not significantly
affect its global structure.
Therefore, throughout this study, we regard solutions in which only a limited
portion interior to the singular point enters region
(\mbox{I\hspace{-1.2pt}V}) as physically acceptable.

To derive the differential equation for $\log_{10} T$,
we combine Eq.~\ref{eq:effdiffusion} with the differential equation for $\log_{10} r$ as a function of $q$.
The remaining variables are required to be constant with respect to $q$.
Therefore, we introduce the following trivial differential equations
\begin{equation}
    \frac{d \log_{10} r_{\rm s}}{dq}
    = 0,
\end{equation}
\begin{equation}
    \frac{d \log_{10} r_{\rm ph}}{dq}
    = 0,
\end{equation}and 
\begin{equation}
    \frac{d \log_{10} \dot{M}}{dq}
    = 0.
\end{equation}

\subsection{The Boundary Conditions}
The number of dependent variables is $6$,
so $6$ boundary conditions are required.
These consist of two conditions at the singular point (Eqs.~\ref{eq:cond_sing1} and \ref{eq:cond_sing2}),
two conditions at the inner boundary (Eqs.~\ref{eq:cond_in1} and \ref{eq:cond_in2}),
and two conditions at the outer boundary (Eqs.~\ref{eq:cond_out1} and \ref{eq:cond_out2}).

\subsection{Iterative Solution Method}
For a mesh with $N$ grid points, the discretized system consists of $6(N-1)$ differential equations.
Together with the boundary conditions, the total number of equations becomes $6N$.
Since each dependent variable is represented by its values at $N$ mesh points, the total number of unknowns is also $6N$.
Therefore, the Jacobian matrix has dimension $6N \times 6N$, and the correction to the solution can be obtained by solving the corresponding linearized system.

The obtained corrections are used to update the solution, and this procedure is repeated until the corrections become sufficiently small.
In the present calculations, we adopt $N = 1000$ grid points, with the singular point located at $N_{\rm s} = 500$.

\section{Formulation of Continuum-driven Solutions} \label{Asec:contonly}
The line-driven wind formulation includes both the continuum radiation force and the line force in a self-consistent manner.
However, this formulation is applicable only to solutions with a positive velocity gradient.
Because the line force depends explicitly on the velocity gradient, the Euler equation cannot be rewritten as an explicit equation for $dv/dr$ and does not admit physically meaningful solutions with a negative velocity gradient.

Such limitations typically arise when the wind region mass is sufficiently large that the gravitational force strongly affects the wind acceleration and a negative velocity gradient appears.

To obtain wind solutions in this parameter regime, we introduce a continuum-driven formulation.
By neglecting the line force, the Euler equation can be rewritten as an explicit equation for the velocity gradient, allowing both positive and negative velocity gradients.
This enables wind solutions to be calculated even in the parameter regime where the line-driven formulation is no longer applicable.

Moreover, these solutions generally have relatively massive wind regions, causing the photosphere to be located farther from the degenerate core.
As a result, the wind acceleration occurs predominantly beneath the photosphere, where the line force is expected to be unimportant.
Therefore, the physically realized solutions in this regime are expected to be well approximated by the continuum-driven solutions.
Finally, the continuum-driven formulation also provides a useful reference for quantifying the contribution of the line force by direct comparison with the corresponding line-driven solutions.

We describe the governing differential equations.
In this model, we consider the region extending from the surface of the degenerate core to the photosphere defined by the Rosseland mean opacity.
The Euler equation includes the gas pressure, radiation pressure, and gravity:
\begin{equation}
    v \frac{dv}{dr}
    + \frac{1}{\rho}\frac{dP}{dr}
    + \frac{GM_{\rm WD}}{r^2}
    =0,
\end{equation}where the total pressure is given by the sum of the ideal gas pressure and the radiation pressure,
\begin{equation}
    P \equiv
    \frac{k_{\rm B}\rho T}{\mu m_{\rm H}}
    +\frac{aT^4}{3}.
\end{equation}The energy transport is described by the diffusion approximation,
\begin{equation}
    \frac{dT}{dr}
    =
    -\frac{3\kappa_{\rm R}\rho L}
    {16\pi acT^3r^2}.
\end{equation}Assuming a steady-state flow, the continuity equation and the energy conservation equation are
\begin{equation}
    4\pi r^2\rho v=\dot{M},
\end{equation}
\begin{equation}
    L+\dot{M}
    \left(
    w+\frac{v^2}{2}
    -\frac{GM_{\rm WD}}{r}
    \right)
    =\Lambda_{\rm tot}.
\end{equation}

We next describe the boundary conditions.
In this formulation, the regularity conditions at the singular point can be written algebraically as
\begin{equation}
    2c_s^2
    -
    \frac{GM_{\rm WD}}{r_{\rm s}}
    -
    \left.
    \frac{d\ln T}{d\ln r}
    \right|_{\rm s}
    \left(
    \frac{k_{\rm B}T_{\rm s}}{\mu m_{\rm H}}
    +
    \frac{4aT_{\rm s}^4}
    {3\rho_{\rm s}}
    \right)
    =0,
\end{equation}
\begin{equation}
    v_{\rm s}^2=c_s^2.
\end{equation}The boundary conditions at the inner boundary are identical to those adopted in the line-driven wind formulation:
\begin{equation}
    r_{\rm in}=R_{\rm WD}(M_{\rm WD}),
\end{equation}
\begin{equation}
    L(r_{\rm in})
    =
    4\pi r_{\rm in}^3
    \rho
    \varepsilon_{\rm CC}.
\end{equation}At the outer boundary, we impose the photospheric boundary conditions,
\begin{equation}
    \kappa_{\rm R}\rho r_{\rm out}
    =
    \frac{8}{3},
\end{equation}
\begin{equation}
    L
    =
    4\pi r_{\rm out}^2
    \sigma T^4.
\end{equation}

The present formulation is almost identical to that of \cite{1994ApJ...437..802K}, except for the inner boundary condition.
Indeed, the governing equations and boundary conditions are essentially the same as those of the line-driven formulation, the only difference being that the Rosseland mean opacity, $\kappa_{\rm R}$, is used instead of the effective opacity, $\kappa_{\rm eff}$.

\bibliography{reference}{}
\bibliographystyle{aasjournalv7}

\end{document}